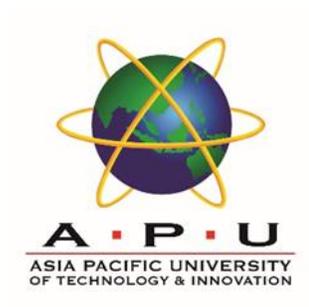

**Optimizing the AI Development Process by Providing the Best Support Environment**

Taha Khamis

DISSERTATION SUBMITED IN FULFILMENT OF THE REQUIRMENTS OF THE DEGREE OF MASTER OF SOFTWARE ENGINEERING

ASIA PACIFIC UNIVERSITY OF TECHNOLOGY & INNOVATION (APU)

JUNE 2022

# ACKNOWLEDGEMENT

      Many scholars, academics, and practitioners helped me with my thesis. I'm grateful for all of their help. They've helped me think about things in a new way. Dr. Homam Mokayed, my main thesis supervisor, has been a constant source of encouragement and support throughout the entire process of preparing this project, from its beginnings to its completion. His unconditional support and outstanding guidance in providing insightful suggestions, direction, expertise, and kindness all contributed enormously to the efficacy in completing this study within the given time frame. In addition, I'd want to thank my second marker Dr. Imran Medi and all of my fellow students for their support and encouragement. This thesis would not be what it is now if it hadn't received the dedication and enthusiasm it has.

# ABSTRACT

The purpose of this study is to investigate the development process for Artificial inelegance (AI) and machine learning (ML) applications in order to provide the best support environment. The main stages of ML are problem understanding, data management, model building, model deployment and maintenance. This project focuses on investigating the data management stage of ML development and its obstacles as it's the most important stage of machine learning development because the accuracy of the end model is relying on the kind of data fed into the model. The biggest obstacle found on this stage was the lack of sufficient data for model learning, especially in the fields where data is confidential. This project aimed to build and develop a framework for researchers and developers that can help solve the lack of sufficient data during data management stage. The framework utilizes several data augmentation techniques that can be used to generate new data from the original dataset which can improve the overall performance of the ML applications by increasing the quantity and quality of available data to feed the model with the best possible data. The framework was built using python language to perform data augmentation using deep learning advancements.

# Table of Contents

# LIST OF FIGURES

viii

# CHAPTER 1
# INTRODUCTION

1.1. Overview

We are living in a world where the machine learning (ML) applications have become an essential part of most of our life's aspects (Kim et al., 2018). Whether we are talking about transportations, healthcare, or any other industrial field (Khalid et al., 2009; Mokayed et al., 2021; Mokayed, Palaiahnakote, et al., 2022) the ML applications have been increasing rapidly (Amershi et al., 2019). The machine learning process is basically a part of what is known as Artificial Inelegance where the data and algorithms are provided to the Artificial Intelligence (AI) in order to make it act and think like a human. The process in which the AI is taught is almost the same as how we learn as humans in our daily lives (Brown, 2021). The successful use of machine learning advancements such as deep learning has changed the world in many ways(Khalid et al., 2011; Ng et al., 2015) and made it impossible not to notice the remarkable applications that have been introduced to the world with the help of machine learning development (Ashmore et al., 2021). The majority of people across the world believe that all of the wonderful technologies brought to the world by the machine learning is just the beginning of this revolution and we should be expecting more sectors to be depending heavily on ML application as the time goes (Makridakis, 2017). The machine learning development can be very different from the traditional development of applications (Brynjolfsson & Mitchell, 2017). The soul difference between traditional programming and machine learning development is that in traditional programming, we are aiming to solve a specific problem by setting a predefined rules or logics as a program while in machine learning the main aim is to feed the AI with a set of algorithms and data to enable it to learn as a human and answering the question or solving the problem by analyzing the input data (Daisy, 2021). It has been reported that while the development of ML applications grows there are also various obstacles that come to light (Nascimento et al., 2019). Understanding the development life cycle of ML apps is a first step in analyzing these problems and attempting to address them. To summarize, a high-level view of the ML application development life cycle can be summarized as follows: first, a problem must be understood; second, data must be collected and preprocessed in a stage called data management; then, a model must be built; and finally, model deployment and maintenance must be performed (Morgunov, 2021).

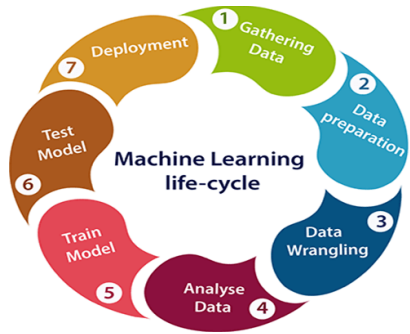

*Figure 1 The machine learning lifecycle*

1.2. Problem Statement

The data management stage (DM) of machine learning development is the most challenging stage, with numerous obstacles arising during this stage (Ashmore et al., 2021). Lack of data sets has a major influence on the model. The most significant issue to the model is the paucity of accessible data. The accuracy of the ML application outputs might be weakened by using datasets that are either too small or incorrectly formatted and organized. A successful start in the early phases of machine learning is crucial for the rest of the process. We must guarantee that the early phases of machine learning, such as data management, are error-free in order to ensure that the final application is accurate. Data augmentation is one of the strategies that can be used by developers to tackle the problems in the DM stage by feeding the model with produced datasets that increase the accuracy of the output as well as fixing the data availability problems. We can increase the likelihood of a more accurate ML application and decrease the amount of time developers spend attempting to address data problems using conventional ways by building a framework that incorporates the most modern technology.

1.3. Research Aim

The primary goal of this project is to improve the artificial intelligence and machine learning development process by providing developers with a framework to utilize throughout the data management stage in order to boost the possibilities of having more accurate and effective model learning results.

1.4. Research Objectives
❖ To analyze and evaluate the impact of insufficient data during data management stage in ML development.

- ❖ To propose the developers with the most advanced techniques to be used in order to solve the issues during data management stage of ML development.
- ❖ To build a framework to be utilized during the data management stage of machine learning development life cycle.
- ❖ To evaluate the impact of using the proposed framework.

## 1.5. Research significance

With a good framework for data management in the machine learning development cycle, the amount of time and effort spent on data collecting and cleaning will be decreased which, according to many studies, is a big issue for developers (Roh et al., 2021). There are several methods and strategies available for dealing with the problems that arise during the data management stage (Wong et al., 2016). As a result, developers may have a hard time deciding which approach or method to apply in any given circumstance. Because of this, when the most cutting-edge and efficient technologies are properly used during the data management stage, developers will save both time and effort while boosting their chances of creating more accurate and successful machine learning apps (Li et al., 2019).

## 1.6. Research Scope

The major objective of this study is to provide the finest support environment for AI development across the development life cycle in order to optimize it. The main focus of this work will be on the data management stage as it's the most critical stage of ML development. There are several issues in dealing with data paucity when it comes to machine learning. This project will provide a framework that addresses both data insufficiency and data accuracy issues. This framework will be evaluated and tested on a particular sort of data (images). For generic ML applications, this framework is expected to enhance accuracy and solve the lack of data issue.

## 1.7. Structure of the Thesis

The remainder of this dissertation is structured in the following manner. The second chapter is a review of the literature. The methodology is described in Chapter 3. The findings and associated discussion along with the implementation are presented in Chapters 4 and 5, respectively. Chapter 6 brings the dissertation to a close and makes recommendations for further research.

# CHAPTER 2
# LITERATURE REVIEW

## 2.1. Introduction

According to a wide range of experts, businessmen, and members of the general public, machine learning will be at the center of many new applications and services in important industries like healthcare(Al Kouzbary et al., 2022; Khamis et al., 2023), transportation, Realtime applications(Mokayed, Quan, et al., 2022), and military in the near future (Ashmore et al., 2021).

It's common for projects to start with an issue that has to be addressed. Tracking the effects of a change will be easier if you have a clear understanding of the issue you're attempting to address (Morgunov, 2021). For the best results from machine learning, you must first describe the issue that you are trying to address (Alake, 2021). Data is soul of any machine learning application (Robert, 2014). After the data management stage completed, a set of datasets will be collected and cleaned to be used in the next stage (Ashmore et al., 2021). The next stage of machine learning development is Model Learning when a model or algorithm is developed from data. All non-functional needs, such as providing outputs in the acceptable time and utilizing a suitable quantity of resources, must be met by a competent model (Ashmore et al., 2021). A full machine learnt model that will be utilized in the following step of machine learning development will be done throughout this stage (von Rueden et al., 2021). There are times when training data sets do not fulfil the model's requirements, and Model Verification guarantees that it meets those criteria even if new data sets come into play (Ashmore et al., 2021). To begin, the trained model serves as the primary input, and a verified model and the verification's results provide enough information to let potential users determine whether the model is acceptable for its intended application. In the last step, the primary input is a verified model and its supporting verification proofs. The major outcome of a well-implemented model is the system's behavior (Hill et al., 2018).

There are various obstacles that arise when the development of machine learning applications grows, according to (Nascimento et al., 2019). Data is at the heart of all machine learning (ML) technologies. A model's inputs and outputs are described by this data, which may be implicit for unsupervised techniques. To put it another way, these data represent the specifications we want to include in our machine learning model. As a result, data must be explicitly considered in every assurance argument (Ashmore et al., 2021).

## 2.2. Data Management

Data is the soul for most of the applications nowadays. Its worth mentioning that keeping our data safe and secure is a critical aspect in all type of applications. A good practice for securing our data could be by using the latest tools for network association risk minimization(Kamalakannan & Khamis, 2022).

Research by Douglass (2020) found that gathering data is the initial step in the data management process. Gathering of data samples for use in a machine learning model by means of observations and measurements is the purpose of data collection (Ros et al., 2016). For example, a student's height or number of pupils or movies seen may be a numerical data point, a category data point, or even plain text like a student's notes or a student's journal (DataRobot, 2021). It is necessary to gather information from many sources in order to create an AI/ML solution that can be put into practice. As a consequence, the data acquired should be relevant to the business issue being addressed. The capacity to gather data allows you to keep a record of earlier occurrences, which may be utilized to find patterns that repeat via the analysis of data acquired. Machine learning algorithms may be used to develop prediction models based on these patterns, which seek for trends and forecast future changes. (DataRobot, 2021). Developing high-quality prediction models necessitates the use of efficient processes for gathering data. The data must be error-free and include necessary information in order to guarantee the task's success.

According to a recent study (Roh et al., 2021), there are three major ways to gather data depending on the scenario. Sharing and searching for new datasets are the primary goals of data gathering. Discover, augment, and produce data sets are some of the methods we might utilize in the data collecting process. According to research by Whang and Lee (2020), obtaining the proper datasets for training models is a large difficulty for ML developers since the amount of data accessible throughout the globe is enormous and constantly expanding, making it difficult for ML developers to discover the suitable datasets. When the data sets are available, the second procedure will be used: data labelling. Individual data may be labelled using a variety of ways, including manual labelling and weak labelling, in this method. If we can make changes to current data rather than creating new datasets, we may apply the last option mentioned above. It is possible to combine these three approaches, according to (Roh et al., 2021).

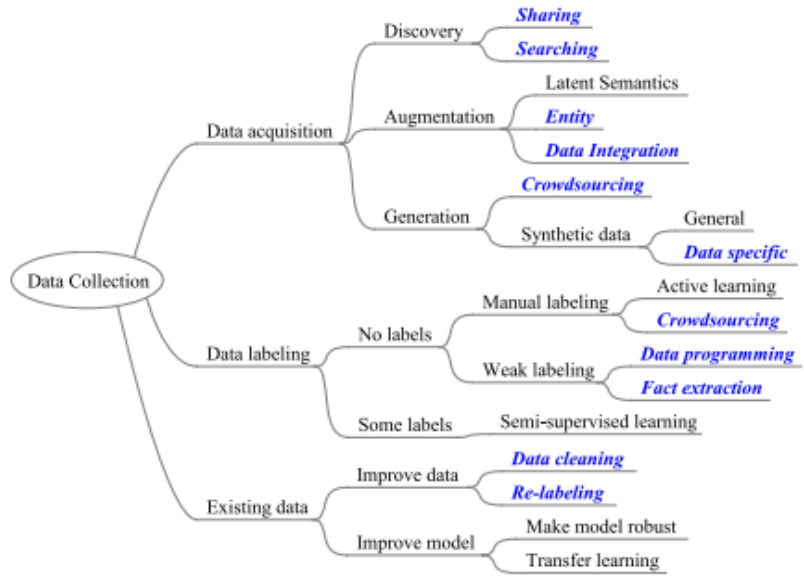

*Figure 2 data collection methods*

We may utilize augmentation techniques if data samples cannot be obtained for any reason or if they are too costly or time-consuming to gather (Ros et al., 2016). Supplementing the acquired data with new samples is accomplished via the use of augmentation techniques (Wong et al., 2016). An additional scenario is when numerous resources' data isn't homogenous or needs preprocessing to maintain consistency in datasets for validation and verification in different sources (Khurana et al., 2018). Aside from reducing the complexity of data, prepressing may also be utilized for training purposes. There may be times when preprocessing is necessary to label data samples for supervised machine learning tasks (Douglass, 2020). Analyzing data allows for the augmentation and preparation of data to be completed (Sharp Sight Labs, 2016).

Following the collection of data sets, a number of actions will be carried out in order to prepare the data for use in the model-building process. Typically, the first action that will be carried out after the acquisition of data is data labelling. Data labelling is the act of classifying and marking data in order to better understand its behavior and trends in order to make better decisions (Borges et al., 2021). When the data sets have been completed, the data labelling process takes over. The semi-supervised strategy is the first approach utilized in data labelling, and it makes use of current labels to predict new labels in the data (Whang & Lee, 2020). Crowdsourcing is an alternate labelling methodology that uses manual labelling and more sophisticated approaches, such as active learning, to finish the labelling process. When implementing crowdsourcing, controlling the quality of processing and post-processing are two of the most difficult issues (Sheng & Zhang, 2019). For certain applications, a small team of in-house information technology professionals might perform better than a big one (Taddeo et al., 2019). Frequently, the person who labels a feature clearly defines the true worth of that characteristic. In other cases, however, this may not be the case: is the cyclist or pedestrian who is riding a bicycle? While ambiguity may be meaningless in certain situations, it is critical in others. Consistencies in labelling are inevitable when the labels are generated by human beings. Finding and correcting discrepancies is an ongoing challenge (Ashmore et al., 2021). Another recent trend is the use of labelling that is too slender. Semi-automatic label production is able to compensate for poor label quality by mass-producing weak labels. In the absence of numerous labels, simpler regulations are preferred. (Whang & Lee, 2020).

Rather of searching for and labelling new data sets in some situations, such as when there is no relevant data to employ in a specific application and no relevant data is accessible, we may improve the quality of present data sets. Instead than wasting time and resources on unneeded data sets, the goal is to make the most of what you currently have (Whang & Lee, 2020). As another example, it is better to clean and/or re-labeled old data rather than add new data sets to the model if the additional data sets do not increase the model's overall accuracy since the data sets are of low quality.

When it comes to the training data, it is extremely common for there to be some inaccuracies in this data (Whang & Lee, 2020). Machine learning systems, such as "TensorFlow Extended," may aid in the process of validation and cleaning by finding problems in data utilizing data visualization and schema construction approaches, as well as other ways (Baylor et al., 2017). Data cleaning is a technique that may be used to rectify inaccuracies in data, and there is a substantial amount of study on the subject (Ilyas & Chu, 2019).

### 2.2.1. Insufficient Data

The process of supplying artificial intelligence with actual data and algorithms in order to enable it to operate like humans while also learning from its own operational experience is referred to as machine learning. In the same manner that everyone of us learns throughout our lives, artificial intelligence is taught in the same way. When it comes to machine learning, however, there is a huge difficulty with the lack of data. A particular quantity of data is required for effective analysis, training, and performance in artificial intelligence; if the data sufficiency is poor, it will not be feasible to complete a dependable project. The scarcity of training data is a critical concern, not only because if AI is unsure about the outcome, it will not express this hesitation, but will instead perform the operation without any alarming indicators. And that's when the difficulty starts to manifest itself. More AI concealed uncertainties and insufficient outputs result from a lack of data. In turn, this leads to the realization that without access, it is impossible to prepare data for machine learning, and even the most exceptional software would be rendered ineffective without adequate data filling. No access also implies that the information is either too difficult to get or does not exist. As a result, enormous potential AI initiatives are often found to be less effective than anticipated, since data scientists are restricted in their ability to acquire and prepare data for machine learning (BroutonLab, 2020).

This is a common issue in the real world. In today's world, sensitive information is stored in a database that only a few people have access to. A firm like HealthCare ensures the privacy of its patients by encrypting their personal information. As a consequence, we will only have a small amount of data from which to make future forecasts, which might lead to inaccurate ones. While healthcare is one of the most prevalent industries, it is far from the only one (Dharshini, 2021).

This is by far the most noticeable barrier when it comes to AI development. When you feed a model with limited amount of data, the model will only provide bad outcomes. Lack of data may express itself in two ways: either as a lack of data or as a lack of excellent data. Most machine learning algorithms need a large amount of data before they can begin to provide useful results. When it comes to this, a neural network is a great example. Neural networks need a vast amount of training data in order to perform effectively. The larger the structure, the more data is needed to get reliable results. Data augmentation might be beneficial to a certain amount, but having more data is always the ideal choice. (Stewart, 2021). According to (Li et al., 2019), the data is a critical player when it comes to determining the chances of having more successful deep neural networks as the effectiveness of neural network rely of the number of data sets that has been fed to the model.

2.2.2. Data Augmentation

Deep learning models, in particular, are highly dependent on the quality, amount, and relevance of training data to perform well. However, lack of data is one of the most significant obstacles to applying machine learning in the workplace. The reason for this is because gathering this information may be expensive and time-consuming in many circumstances. It's possible to speed up machine learning model development by supplementing training data with other sources of data (Takimoglu, 2022).

Artificially increasing the quantity of data by creating additional data points from existing data is known as data augmentation. Using deep learning models to produce new data points or making modest modifications to data are examples of data augmentation. Artificial intelligence (AI) applications, particularly in the deep learning area, are continuing to diversify and grow at a fast pace. Techniques for data augmentation may be a useful weapon in the fight against the issues that the artificial intelligence field is facing. In order to enhance the performance and outputs of machine learning models, data augmentation may be used to provide fresh and varied instances to be used in training datasets. It is more accurate and more efficient for a machine learning model to perform well if the dataset it uses is rich and robust. In the case of machine learning models, the process of gathering and categorizing data may be time-consuming and expensive. Companies may lower these operating expenses by transforming datasets and using data augmentation methods to improve their efficiency. Cleaning data is one of the processes in the creation of a data model, and it is essential for models with high accuracy. However, if cleaning data affects the representability of the data, the model will be unable to make accurate predictions for real-world data inputs. Machine learning models become more resilient as a result of data augmentation strategies, which provide variables that the model may encounter in real-world situations (Takimoglu, 2022).

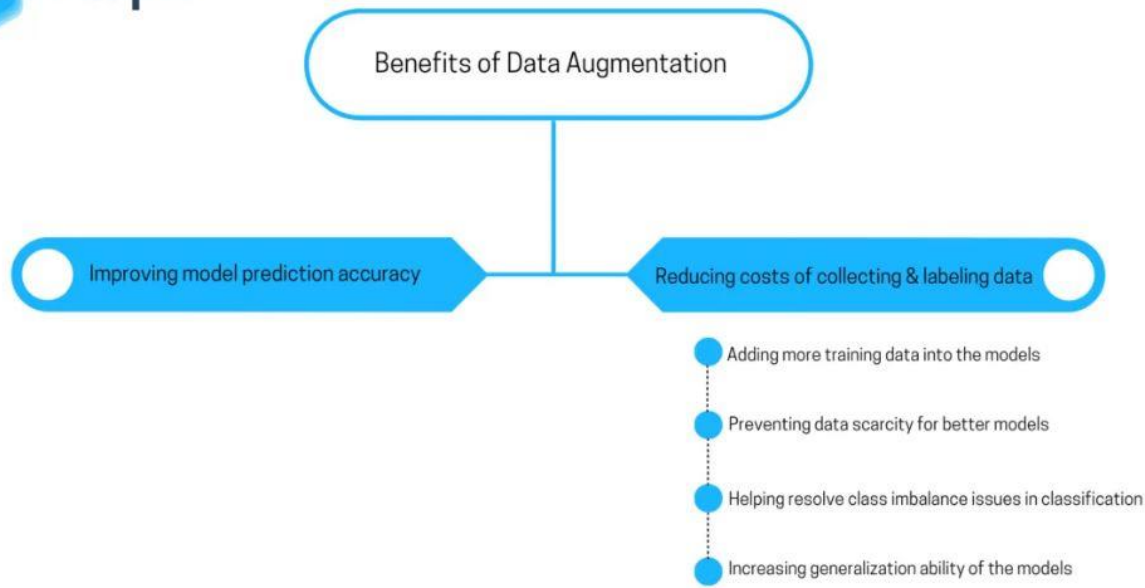

*Figure 3 The Benefits of Data Augmentation*

Convolutional neural networks (CNNs) have become the most often used tool for image analysis and classification as a result of the growing interest in deep learning techniques. In a range of classification tasks, CNNs produced state-of-the-art outcomes; yet, despite their broad perspectives, they still face a number of obstacles to overcome. In large part, this is due to the vast scale of the networks, which may approach millions of parameters, along with a scarcity of trustworthy training data sets. They also have issues with overfitting and lack of generalization skills. Last but not least, avoiding adversarial threats that might cause DNNs to misinterpret data is becoming a big source of worry for experts. In order to overcome these difficulties and get better outcomes, the researchers are working on altering the network design, implementing new learning algorithms, and collecting data. The most often encountered issue is a lack of high-quality data or an imbalanced class balance within the datasets. These days, the most efficient DNNs are quite massive, and as a result, they demand a great quantity of data, which in many circumstances may be difficult to get. For example, the extremely famous CNN architecture VGG16 is composed of 16 layers of neurons and has a total of 138 million of parameters in its entire configuration. Furthermore, the effectiveness of new designs is often evaluated using ImageNet, a dataset that contains more than one million pictures from 1000 non-overlapping categories. This challenge may be addressed in a number of ways, including the use of data augmentation and data synthesis. With the increasing popularity of CNNs, there has been a tremendous growth in interest in data augmentation techniques. To perform traditional affine and elastic transformations on data, which have been shown to be effective, is the most popular and widely used current practice for data augmentation. Examples include creating new images by performing rotation or reflection of the original image; zooming in and out; shifting; applying distortion; and changing the color palette. Despite their many benefits, basic classical procedures are not always sufficient to considerably increase the accuracy of neural networks or to address the issue of overfitting. Furthermore, recent research into so-called adversarial assaults on CNNs has shown that deep neural networks may be readily misled into misclassification of pictures simply by performing partial rotations and image translations on the images being trained on. Adding noise to photographs and even modifying one, carefully picked pixel in an image are all possible options. Increased dataset size, achieved by data augmentation and picture synthesis, makes the dataset more resilient and less subject to adversarial assaults (Mikolajczyk & Grochowski, 2018).

Existing benchmarks may be improved by applying deep convolutional networks to Computer Vision applications, as shown by a number of ongoing studies (Shorten & Khoshgoftaar, 2019). One of the most challenging aspects of these models is to increase their ability to generalize. A model's "generalizability" refers to how well it performs when tested on previously observed data (training data), as opposed to how well it performs when tested on data that has never been seen before (testing data). Models with poor generalizability have overfitted training data. Overfitting may be discovered in part by plotting the overfitting data. At each epoch of the training and validation procedure, the accuracy is maintained. Visual depiction of overfitting based on these accuracies over training epochs is shown in the graph below.

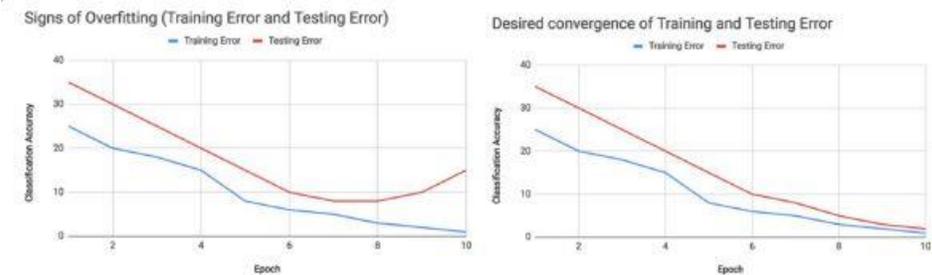

*Figure 4  As the training rate decreases, the left plot indicates an infection point where the validation error begins to grow. Overfitting to the training data has resulted in a model that performs badly on the testing data.*

Deep Learning models need decreasing validation errors in order to be useful. due to a blunder in training Data augmentation is a powerful tool when it comes to doing this. The enhanced data will reflect a more comprehensive set of possible data points. the distance between the training and validation sets as well as any other possible issues to be incorporated in future testing sets should be minimized (Shorten & Khoshgoftaar, 2019). Using Data Augmentation approaches, which go above and beyond what's previously been done, the training dataset is the beginning point. Based on the idea that augmentations may extract more information from the original dataset, this approach is taken. In Tese augmentations, the size of the training dataset is lowered artificially by data warping or oversampling. The original label of a picture is preserved when using data warping augmentations. Adversarial training and neural style transfer are among the many innovations that fall under this category. A technique called oversampling augments the training set by creating synthetic examples. Images are mixed, feature spaces are augmented, and generative adversarial networks are all part of the process here (GANs).

More data leads to improved Deep Learning models, as is widely acknowledged (Sun et al., 2017). To put it another way, the human effort of gathering and classifying large datasets may be a daunting endeavor. Medical image analysis has the unique problem of working with little datasets. Given large data, deep convolutional networks have been found to be particularly strong for medical image processing tasks such as skin lesion classification (Esteva et al., 2017). As a result, CNNs are now being used for a variety of medical image processing tasks (Litjens et al., 2017), including the classification of liver lesions, brain scans, and skin lesion classification. MRI and CT scans, both of which require a lot of time and money to obtain, provide a lot of the pictures used in this research. Developing large medical image databases is difficult owing to the rarity of illnesses, privacy, the necessity for medical specialists to classify, and the cost and physical effort required to undertake medical imaging operations. Oversampling using GANs has been the focus of numerous research in the medical picture classification field because of these challenges (Shorten & Khoshgoftaar, 2019). Popular academic picture datasets are often used to assess the efficacy of Data Augmentation research. It includes CIFAR-10/100 and ImageNet, as well as SVHN (street view house numbers), Caltech-101/256, MIT locations, MIT-Adobe 5K, Pascal VOC, and Stanford "Cars datasets. CIFAR-10, CIFAR-100", and ImageNet are the most commonly mentioned datasets. Open-source dataset enlargement has provided academics with a broad range of examples to compare the outcomes of Data Augmentation approaches. Big data is a term used to describe databases such as ImageNet. In order to imitate restricted data, many studies confine themselves to a small subset of the dataset. Imagination and daydreaming are both examples of data augmentation. Humans are capable of imagining a wide range of possibilities depending on their own experiences. We can learn a lot about the world around us if we use our imaginations. As a result of data augmentation techniques such as GANs and Neural Style Transfer, computers can better comprehend pictures by 'imagining' changes to them (Shorten & Khoshgoftaar, 2019).

### 2.2.3. Image Augmentation

When trying to understand why picture recognition is so tough, it's important to examine the context in which the problem arises first. It's very uncommon for image recognition software to face challenges like distinguishing between a cat and a dog due to differences in perspective or occlusion (the backdrop). For Data Augmentation to work, you must include these translational invariances into your dataset in such a manner that your final models still perform well despite these challenges.

#### 2.2.3.1. Image Augmentation Background

LeNet-5 (Haykin & Kosko, 2001) makes advantage of data warping to improve images. Using CNNs to classify handwritten numbers was one of the first and most effective uses of this technology. Besides oversampling, data augmentation has been studied in a variety of industries. This method of resampling unbalanced class distributions can be used to guarantee that the model is not unreasonably biassed towards identifying examples as belonging to the majority class type. For class ratio optimization, random sampling (ROS) repeats minority class photos until the desired ratio is obtained. SMOTE (Synthetic Minority Over-sampling Technique) was created by Chawla et al. to develop oversampling approaches (2002). New points are interpolated from current SMOTE and the Borderline-SMOTE variations using k-Nearest Neighbors, an extension of SMOTE (Fan et al., 2011). Because of class imbalance, SMOTE was frequently applied to both tabular and vector data sets.

Using convolutional networks, the AlexNet CNN architecture (Krizhevsky et al., 2017) revolutionized image classification and classification approaches.

Their research makes use of Data Augmentation, which increases the dataset's size by a factor of 2048, a considerable increase. This is achieved, among other things, by the use of PCA color augmentation and the random clipping of 224*224 patches from the source pictures. This Data

Augmentation approach helps to reduce overfitting while training a deep neural network. The authors claim that the model's error rate was reduced by more than one percent as a result of their augmentations.

Neural style transfer and neural architecture search have all been released since then, with GANs being the first of the three in 2014, according to Goodfellow, Gatys and Gatys et al. (Gatys and Gatys et al., 2015). Papers such as those that describe DCGANs, CycleGANs, and Progressively-growing GANs, as well as their expansions, were published in 2015, 2017, and 2018. (Karras et al., 2018). Creating Perceptual Losses (Johnson et al., 2016) resulted in an increased rate of neural style transfer, as seen in this study. Neural Augmentation (Perez & Wang, 2017), Smart Augmentation (Lemley et al., 2017), and AutoAugment (Lemley et al., 2017) are current examples of how meta learning techniques from NAS may be used to Data Augmentation. These studies were published in 2017, 2017, and 2018, respectively.

Deep Learning in medical imaging has been a key usage of CNNs since they were first revealed in 2012. There has been a rise in interest in the domains of deep learning and medical imaging after (Esteva et al., 2017) shown that they can identify skin cancer at the level of a dermatologist using deep learning and medical imaging.

(Yi et al., 2019) that GANs have a long history of successful application in medical imaging, according to their findings. In recent years, GANs have been used in object reconstruction, including CT denoising, MRI acceleration, and PET denoising, and the use of super-resolution GANs in the segmentation of retinal vasculature (Mahapatra & Bozorgtabar, 2017). Additionally, other work by (Yi et al., 2019) discusses the use of GAN image synthesis in medical imaging applications like brain MRI reconstruction (Plassard et al., 2018), lung cancer detection (Chuquicusma et al., 2018), high-resolution skin lesion reconstruction (Baur et al., 2018), and classification of chest x-ray abnormalities (Yi et al., 2019). (Yi et al., 2019). As of 2018, (Moradi et al. The GANs may be used to generate images. Data augmentation was utilized to classify liver lesions (Frid-Adar et al., 2018). Using GAN-based Data Augmentation, classification performance was increased by 85.7 percent in sensitivity and 92.4 percent in specificity, resulting in a classification score of 76.6 percent in sensitivity and 84.4 percent in specificity, respectively.

Most of the augmentations that have been tested so far seem to be based on image recognition. It is a model that utilizes a picture to predict an output label, such "dog" or "cat," from an input image.

Image recognition may help with many Computer Vision applications, such as detecting objects. YOLO, R-CNN, fast R-CNN (Redmon et al., 2016), and faster R-CNN (Ren et al., 2017) are a few examples, as are approaches like U-Net (Weng & Zhu, 2021) and Fast R-CNN (Ren et al., 2017). These are just few examples (Redmon et al., 2016). As of 2017, (Shelhamer and co-authors).

2.2.3.2.   Image Data Augmentation techniques

There are two basic categories of image augmentation techniques: white-box and black-box methods based on (DNN) deep neural networks. Each has its own set of benefits (Mikolajczyk & Grochowski, 2018). Data For many years it was thought that simple changes like horizontal flipping or color space augmentations were all that was needed for augmentation to work. Numerous invariances are encoded by these alterations. Detailed explanations of how each augmenting approach works, as well as the outcomes of experiments, are provided in this section.

2.2.3.3.   'Data Augmentations based on basic image manipulations'
- ❖ 'Geometric transformations'

This section explains a variety of image processing techniques, including geometric transformations and many more. One way to categorize the types of augmentations presented in the next section is according to how simple they are to perform.

- ❖ Flipping

Flipping the horizontal axis is far more common than flipping the vertical axis. CIFAR-10 and ImageNet datasets have proved the benefits of this simple strategy, which is easy to apply. For datasets that need text recognition, such as MNIST or SVHN, this is not an appropriate modification (Shorten & Khoshgoftaar, 2019).

- ❖ Color Space

Tensors of the dimension (height, width, and color channels) are often used to encode digital picture data. Another method that is easy to execute is to do augmentations in the space of the color channels. 'Isolating a single-color channel such as R, G, or B' is a simple way to enhance a picture's color. By removing the other color channels' zero matrices and adding them to the isolated matrix, a picture may be swiftly transformed into a representation in a single-color channel. With basic matrix operations, the brightness of a picture may be increased or decreased by manipulating RGB values. Deriving a color histogram is the first step to more complicated color augmentations. In picture editing software, adjusting the intensity levels in these histograms alters the lighting (Shorten & Khoshgoftaar, 2019).

❖ Cropping

Images with mixed height and width measurements may be processed by simply cropping the center portion of each picture. The random cropping technique is also capable of producing results that are very comparable to those obtained by translations. While translations retain spatial dimensions, random cropping reduces the size of a picture by a factor of two, such as from (256,256) to (224,224). This may or may not be a label-preserving transformation depending on the cropping threshold (Shorten & Khoshgoftaar, 2019).

❖ Rotation

It is possible to add rotational augmentations to a picture by rotating it on an axis between 0° and 359°. Depending on the degree of rotation, the safety of rotational augmentations may be greatly affected. On digit identification tests like MNIST, small rotations like those between 1 and 20 or 1 to 20 may be helpful, 'but as the rotation degree grows, the label of the data is lost after transformation' (Shorten & Khoshgoftaar, 2019).

❖ Translation

To eliminate data bias due to picture location, images may be shifted left, right, up, or down. A face-recognition model would, for example, need to be evaluated on precisely centered photos if all of the images in a collection are centered. A random or Gaussian noise may be used to fill the empty area left after translating the source picture in one way or another. This padding keeps the image's post-augmentation spatial dimensions intact (Shorten & Khoshgoftaar, 2019).

❖ Noise injection

It is possible to introduce a Gaussian distribution-derived random matrix into a system by way of noise injection. Using nine datasets from the UCLA repository, (Moreno-Barea et al., 2019) analyzed the impact of noise introduction on the datasets (UCI Machine Learning Repository, n.d.). Increased noise in photos may help CNNs learn more robust characteristics. Geometric adjustments may successfully compensate for positional biases in the training data set. As a result of such biases, it is possible that training data is distributed differently from testing data. Because each face is exactly centered in the picture, geometric alterations are an excellent way to eliminate positional bias in a dataset of faces for facial recognition. Additionally, geometric modifications have the advantage of being simple to do while at the same time being effective in countering positional bias. A number of image processing libraries are available that make basic operations like horizontal flipping and rotation a breeze. Geometric modifications include additional drawbacks, such as the need for more memory, higher processing, and a longer learning period. The label of the image must be manually inspected after geometric alterations, such as translation or random cropping, in order to ensure that it has not been changed in any way. To sum it all up, biases are more complex in many of the stated application fields, such as medical image analysis, than the ordinary positional and translational alterations. Therefore, geometric changes can only be used in a limited number of locations and at restricted periods (Shorten & Khoshgoftaar, 2019).

- ❖ Changing the color space

There are three identically-sized matrices used to encode the data, each of which comprises a picture. Tese matrices are used to express RGB color values. When it comes to the process of recognizing images, one of the most common stumbling blocks is bias in the lighting. As a result, the efficiency of color space conversions, sometimes referred to as photometric transformations, is easy to grasp. To quickly adjust photographs that are too light or dark, cycle over the images and change the pixel values by a fixed amount. Color space alteration may also be done quickly by cutting off individual RGB matrixes. Pixel values may also be restricted to a certain minimum or maximum value as another transformation. Color in digital photographs may be augmented in a variety of ways because of its inherent representation. It is also possible to create color space changes using image-editing programs. A color histogram is formed by aggregating the pixel values in each RGB color channel of a picture. Filters that alter an image's color space may be applied using this histogram's control points. The RGB matrices in picture databases may be reduced to a single grayscale image for easier representation. Since height width 1 results in smaller pictures, quicker calculation is possible. But this has been found to have a negative impact on performance. On ImageNet (Deng et al., 2010) and (Everingham et al., 2015), (Chatfield et al., 2014) observed that grayscale and RGB pictures had a 3% decline in classification accuracy. Digital color may be represented in a variety of ways, including RGB vs grayscale pictures and HSV (Hue, Saturation, and Value). Many color space representations are examined by (Jurio et al., 2010), including RGB, YUV, CMY and HSV.

Color space transformations have some of the same drawbacks as geometric transformations, such as memory, cost, and training time. The labelling of certain colors is lost in the process of transforming them, therefore This is not necessarily a transition that preserves the original label. In the event that the pixels of a picture are reduced to imitate a darker environment, for example, the objects in the image may no longer be visible. Image Sentiment Analysis (You et al., 2015) is another example of a non-label preserving color modification. Using CNNs, an image's sentiment score may be graphically predicted, with scores ranging from "very negative" to "extremely negative," "neutral," or "positive." The presence of blood is an indicative of a bad or very unfavorable impression. Blood may be distinguished from water or paint by its intense red hue. Image Sentiment Analysis will be hampered if the model is unable to distinguish between crimson blood and green paint as a result of frequent color space transformations. It is possible to alter the dataset to emphasis spatial attributes rather than colors by using color space transformations. Color, on the other hand, is a crucial distinguishing element for various jobs (Shorten & Khoshgoftaar, 2019).

- ❖ Kernel filters

Kernel filters can be used to sharpen or blur images. Tese filters work like this: They either apply Filter with a significant contrast between the top and bottom edges or a Gaussian blur to a photograph, which enhances the clarity of the image's edges. They operate in this manner. Images that are blurred may be more resistant to motion blur when tested for Data Augmentation. Additionally, by sharpening photographs for Data Augmentation, you may be able to get more information about the subject matter being photographed. There are two traditional ways to apply kernel filters: sharpening and blurring. This kernel filter (Kang et al., 2017) randomly swaps the values of the pixels within a Nn-sliding window. This technique is known as PatchShufe Regularization. In comparison to the 6.33 percent error rate obtained without PatchShufe Regularization, a 5.66 percent error rate was achieved with PatchShufe Regularization on CIFAR-10. Various filter widths and probabilities of shuffled pixels were examined at each stage of the experiment. To make this work, I used hyperparameters with a 0.05 probability of switching and 22 filter combinations. The Tese tests were conducted using the ResNet (He et al., 2016) CNN architecture (Figs. 5, 6). Data Augmentation's kernel filters are an unexplored frontier. This technique's internal mechanics are quite similar to CNN's internal operations. Using parametric kernels, CNNs identify the best method to represent pictures layer by layer. It is possible to apply PatchShufe Regularization using a convolution layer, for example. Traditional convolution layer settings may be modified to retain spatial resolution and keep subsequent activation layers mapping pixels to values between 0 and 1, as opposed to using the standard activation sigmoid that maps pixels to values fall between 0 and 1 for each layer.

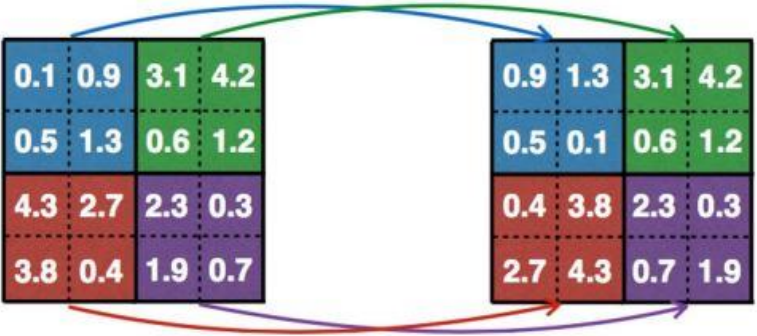

*Figure 5' examples of applying the PatchShufe regularization technique'*

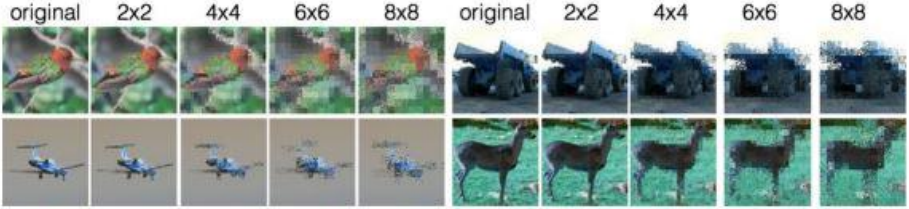

*Figure 6 'Pixels in a n×n window are randomly shifted with a probability parameter p'*

Since kernel filters are used in the network rather than as a supplement to the dataset, it is preferable.

- ❖ Mixing images

When it comes to Data Augmentation, it's paradoxical to use pixel values as a means of mixing pictures together. A human viewer will not be fooled by the visuals that are created as a result of this process. The pairing of samples by (Inoue, 2018) was shown to be an effective enhancement approach. A 224*224-pixel cut and a horizontal flip are performed at random on two photos in this experiment. The RGB channels' pixel values are averaged to create a new picture. Classification models may be trained using this jumbled picture. The new picture has the same label as the first one that was chosen at random (Fig. 7)

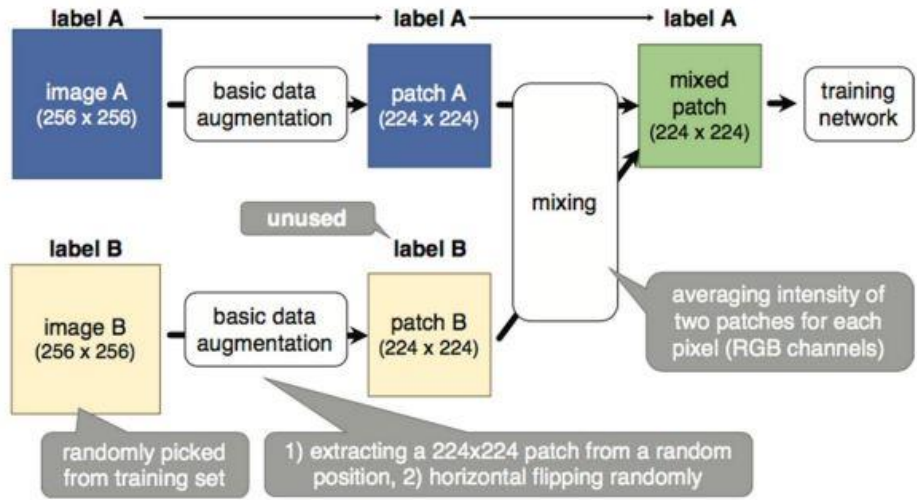

*Figure 7 Sample Pairing augmentation strategy*

- ❖ 'Random erasing'

Another key Data Enhancement method introduced by is random erasure (Zhong et al., 2020). However, random erases occur in the input data space rather than the network architecture, therefore there are some parallels between dropout regularization and these random erases. The primary goal of this strategy was to eliminate problems with image identification caused by occlusion. When an item is occluded, a portion of it is hidden. Because it forces the model to acquire new descriptive features, the random erasing method prevents the model from becoming too dependent on a single visual component. In order to ensure that a network doesn't merely focus on a small region of the picture, random erasure might be a viable option. In order to erase a random portion of the picture, the algorithm randomly selects a portion of the image and applies masking to it. This resulted in a decrease in the CIFAR-10 error rate from 5.17 to 4.31 percent. Random values were proven to be the best way for filling in the patches. When it comes to implementation, you'll just need to create the fill method and mask size by hand (Figs. 8) (Shorten & Khoshgoftaar, 2019).

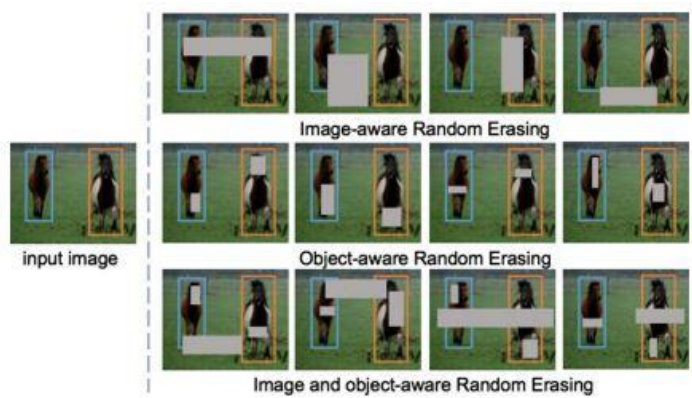

*'Figure 8 Example of random erasing on object detection tasks'*

2.2.3.4. Deep learning-based data augmentations
- ❖ Enhancement of feature spaces

The images in the input field are used for all of the augmentation methods discussed above. Transforming high-dimensional inputs into low-dimensional representations is a specialty of neural networks. Tese networks, which can map images to binary classifications or n1 vectors in thicker layers, can be used to classify images and extract relevant information. There are techniques for removing intermediate representations from neural networks by altering neural networks' sequential processing. Lower-dimensional representations of visual data can be found and retrieved from fully-connected layers (Konno & Iwazume, 2018).

Auto-encoders are very beneficial for doing feature space augmentations on data. Encoding pictures into low-dimensional vector representations and then decoding these low-dimensional vectors back into the original image is how autoencoders function. The encoded representation is utilized to enhance feature space (Shorten & Khoshgoftaar, 2019). It is possible to use auto-encoders to apply feature space enhancements and then rebuild the new instances into input space. Only by extracting vector representations from a CNN can feature space augmentation be performed. Slicing the output layer of the network and replacing the class label with a low-dimensional vector is how this is accomplished. In order to construct vector representations, the training data is first routed through a truncated CNN. Naive Bayes, Support Vector Machines, and even a fully-connected multilayer network may all be trained using these vector-based representations. In order to determine the method's effectiveness, more research is needed.

It may be tough to comprehend the vector data while utilizing feature space augmentation. To transform the generated vectors into pictures, an auto-encoder network may be employed. However, this entails recreating the whole encoding portion of the trained CNN in its entirety. An auto-encoder is a sophisticated and time-consuming tool for training deep CNNs. Data-space augmentation trumps feature-space augmentation when pictures can be translated into data space (Wong et al., 2016).

- ❖ "Adversarial training"

Adversarial training is a technique for investigating the potential of prosthetics. Adversarial training refers to training two or more networks having conflicting objectives encoded in their loss functions. An adversarial attack technique and adversarial training will be examined in this section. Adversarial attacks may be launched by instructing a rival network to make changes to images that cause mistakes in the system's classifications. The adversarial network has been surprised by the effectiveness of these noise-injection-based adversarial assaults. Because it goes against everything, we've learned about how these models represent pictures, this is a huge shock. The adversarial assaults show that representations of pictures are much less resilient than previously thought. As (Moosavi-Dezfooli et al., 2016) show, Deep Fool, a network that finds the smallest amount of noise injection necessary to create a misclassification with high confidence, demonstrates this effectively. It is possible to misclassify 70.97 percent of photos by altering only one pixel, according to (Su et al., 2019). With adversarial assaults restricted to the borders of pictures, (Zajac et al., 2019) produce misclassifications. As picture resolution grows, the success of hostile assaults is magnified. Targeted or untargeted attacks by an adversarial network may refer to the deliberate attempt of the network to induce classification errors. Adversarial assaults are more effective than typical classification metrics in revealing weak decision boundaries. Accurately looking for augmentations may be done by adversarial training in addition to assessment metrics and protection against adversarial assaults. An adversarial network may learn to create augmentations that lead to misclassifications by restricting the set of augmentations and distortions accessible to it. Tese augmentations may be used to strengthen the model's weak points. For this reason, adversarial training might be a useful search strategy for data enrichment. This is in stark contrast to the earlier, more conventional methods of augmentation. However, even if the adversarial augmentations don't replicate real-world situations, they may help strengthen the weak points in the learnt decision boundary that do. Deep CNN models have been shown by (Logan Engstrom, Ludwig Schmidt, Dimitris Tsipras, 2018) to be susceptible to errors introduced by simple transformations such as rotations and translations. ImageNet (Top 1), CIFAR10 (by 72%), and MNIST (by 26%) were all affected by the worst random modifications. Using adversarial examples, (Goodfellow et al., 2015) enhance the MNIST classification problem. With an average confidence of 97.6 percent, a maxout network (Goodfellow et al., 2013) misclassified 89.4 percent of the adversarial instances generated using the "rapid gradient sign method" methodology.

Comparing performance on original testing data with adversarial cases is done by (Li et al., 2018). Contrarian training, whether it takes the form of noise or augmentation search, has yet to be well evaluated and understood for its effectiveness. It has been proven that using adversarial search to inject noise improves speed, however it is not apparent whether this is helpful in decreasing overfitting.

- ❖ GAN-based Data Augmentation

Another intriguing technique is the application of generative modelling for Data Augmentation. A dataset is used to generate new instances, and these new examples are then compared to the original set to see whether they have any traits in common (Shorten & Khoshgoftaar, 2019). GANs, a fascinating and widely used generative modelling framework, were developed as a result of the adversarial training concepts described above. For example, GANs may be used to extract new information from a dataset, according to Bowlles et al. (2018). Despite the fact that GANs aren't the only generative modelling technique out there, they have a substantial advantage in terms of computation speed and the quality of their outputs.

Additionally, there is an intriguing approach for generative modelling that is worth noting: variational auto-encoders (VAE). Variational auto-encoders' sample quality can be improved by expanding the GAN framework (Doersch, 2016). By testing with a variety of inputs, variational auto-encoders develop a low-dimensional representation of the data. An image tensor that was previously described as having size HWCHLC channels is now represented as an image vector with a size n1, which is the same as before in terms of feature space augmentation. Representation in vector representations with low-dimensional constraints will be inferior, even if these constraints are better for visualization when using techniques such as T-SNE (Van Der Maaten & Hinton, 2008). Consider an autoencoder-generated vector representation with a size of 51. It is possible for those autoencoders to map a distribution of labelled data into this space. It is possible to build classes like "head turned left," "centered head," or "headed left," depending on the user's preference. If you utilize vector operations such as adding and subtracting, the encoder may simulate a front view-3D rotation of a new instance using the low-dimensional representation learned by the auto-encoder. A GAN's performance can be considerably improved by using varying auto-encoder outputs (Goodfellow et al., 2014). Bidirectional GANs (Donahue et al., 2017) may also be used to conduct vector manipulation on GAN noise vector inputs.

Recently, there has been a lot of discussion in applying GANs to the problem of data augmentation because of GANs' impressive performance. Additional training data may be generated by these networks in order to improve the performance of classification models. First established by Ian Goodfellow, the GAN architecture is a paradigm for generative modelling based on adversarial training (Goodfellow et al, 2014). The best approach to understand GANs is to compare a police officer to a counterfeiter. The counterfeiter needs some kind of input to work with (generator network). Random vectors to another image to text and many more alternatives are possible here. Discriminators (police) cannot tell if money is real or counterfeit because counterfeiters have learned how to make money in such an inconspicuous fashion that they can't be detected. False or real depends on whether or not an example from the training set was used to generate it or was generated by the generator network.

GANs in network intrusion detection can be better understood by making the analogy between a counterfeiter and a thief. It was possible to fool a black-box detection system using a generator network (Lin et al., 2018). One of the most fascinating aspects of GANs is their capacity for learning. Minimax strategy and Nash Equilibrium (Fedus et al., 2018) game theory analysis methodologies indicate that the generator will trick the discriminator and trick the discriminator. Generative modelling applications benefit greatly from the generator's capacity to surpass the discriminator. Generative adversarial networks (GANs) are the most promising data augmentation generative modelling tool (Shorten & Khoshgoftaar, 2019).

2.3. Conclusion

There are a large number of techniques and methods available for the use of data augmentation. the decision making on which method to use very much depends on the situation and the available resources at hand. In many scenarios its proven that the traditional methods of data augmentation can have a successful impact on the end model by increasing the accuracy of the model or by simply compensate the insufficient data. The more advanced techniques also had proven their success in different situations, however, its more costly and requires high qualified technical people in order to get the most out of it.

# CHAPTER 3
# Methodology

## 3.1. Introduction

Data Management stage is indeed a critical stage in machine learning development because of the heavily dependency of machine learning applications on data (Ashmore et al., 2021). The time and effort spent on this stage is a real challenge because this sage consists of many activities and sub activities in which the whole application success relies on this stage. If the data given to the machine learning application is insufficient or not clean then no matter how good is the end model, it will not perform as expected (Robert, 2014). Data augmentation is one of the main keys to ensure the success of the project by supplying the machine learning application with enough data to perform its intended use. Its proven that data augmentation not only solve the lack of data in some scenarios but also in the situations where the data is available, by utilizing the data augmentation methods, the accuracy of the model is improved quite much (Ashmore et al., 2021).

## 3.2. Research philosophy

This project was aiming to reduce the time and effort spent on data management stage by creating a framework which can be used to perform the augmentation techniques in the shortest time possible without having to write any piece of code by just specifying the path of the data at hand which will then be uploaded to the application and the developer/researcher choose the techniques of augmentation that they wish to perform from the list on the GUI of the application.

The framework then will execute the augmentation methods chosen and will give the images a proper label to reduce the time spent on labelling the data s well as classifying them into their respective folders. The following flowchart describes the method of this research and how its conducted (fig 9).

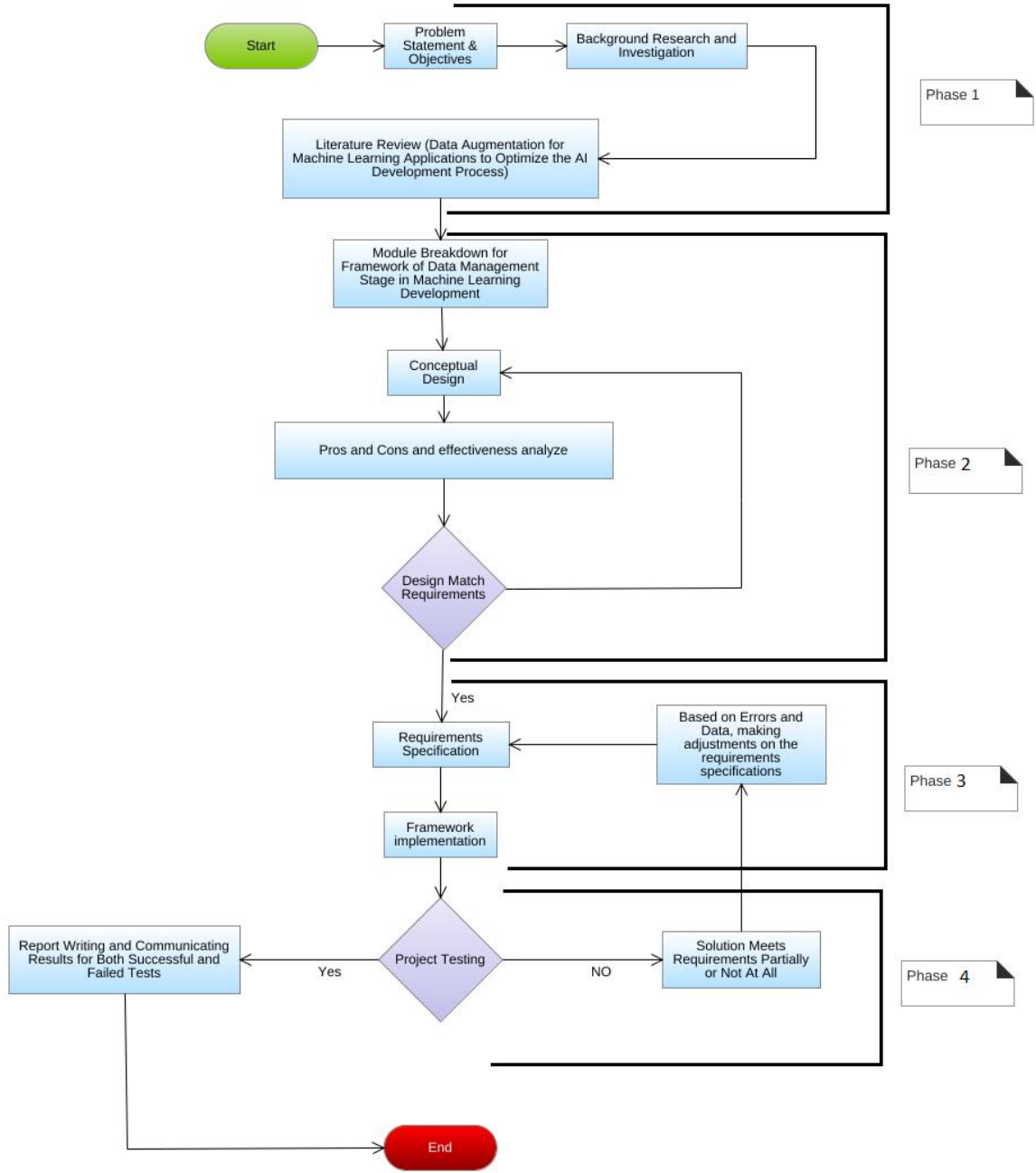

*Figure 9 Research methodology*

Figure 1 depicts the four steps of the approach. Listed below are descriptions of each phase.

3.2.1. Gaining knowledge and understanding the context.

Comprehensive evaluation of data management stage on machine learning development life cycle, the issues encountered, and solutions used in the literature. Analysis of the principal limits and possible issues is then carried out. This phase's primary objective is to get an understanding of current developments in machine learning.

3.2.2. The initialization process of creating a data management framework (Conceptual design).

On the basis of prior stage, a breakdown of modules encompassing requirements, activities and programming designs will be presented according to the scale of the needed system. Drawings of the flowchart, activity diagram, and necessary designs will be specified, and the programming needed for the functions will be produced.

3.2.3. The beginning of the process of creating a data management framework (Development).

During phase 2, the criteria for the idea will be finalized. After verifying that the ideas adhere to the specifications, the framework will be ready for utilization.

3.2.4. Testing and Evaluation

The final application will be evaluated based on the results of phases 1 and 2. In order to improve the framework and address potential difficulties, the data obtained may be utilized. The framework's effectiveness and correctness will be assessed primarily.

3.2.5. Research techniques
- ❖ This study was carried out utilizing quantitative methods. This research investigates the link between machine learning accuracy and the use of data augmentation solutions.
- ❖ Experiments will be conducted to test the framework's ability to improve data management.

3.2.6. Research plan

| Activity | Duration | Start Date | End Date | Week 1-13 | Comments |
|---|---|---|---|---|---|
| Problem statement, objectives, and literature review | 3 weeks | 17-Jan-22 | 7-Feb-22 | 1-3 | Review and final approval |
| Module breakdown and gathering information about the resources. | 1 week | 7-Feb-22 | 14-Feb-22 | 4 | Identifying the activities and sub activities |
| Conceptual design and design decision | 2 weeks | 14-Feb-22 | 28-Feb-22 | 5-6 | Different appoaches to answer the questions |
| Gathering resources | 1 week | 28-Feb-22 | 7-Mar-22 | 7 | |
| Development of framework | 3 weeks | 7-Mar-22 | 28-Mar-22 | 8-10 | Developing the framework |
| Framework testing | 2 weeks | 28-Mar-22 | 11-Apr-22 | 11-12 | Testing the framework using multiple scenarios |
| Report finlizing and submission | 1 week | 11-Apr-22 | 18-Apr-22 | 13 | Reporting the result of the filed testing and the framework |

*Figure 10 Gantt chart of Activities.*

## 3.3. Research Approach

The approach of this research is to combine multiple data augmentation techniques that have proven their success on improving the accuracy of the model and solving the insufficiency of data. The combined techniques are as follow:

- Rotate
- Shift
- Shear
- Zoom
- Flip
- Brightness change
- Noise injection

Besides augmenting new data, the application will be able to label the new augmented data with its respective labels and sort the data in different files.

3.3.1. Rotating augmentation

When it comes to image rotating, the framework will perform image rotation in random degrees and generates 7 new images from the original source. Random Rotate is a valuable augmentation in particular because it alters the angles at which items appear in your dataset during training, which is good for improving accuracy. Perhaps photos of an item were only captured horizontally throughout the image collecting phase, but during the manufacturing process, the thing might be skewed in any way. Random rotation may help you enhance your model without the need to gather and identify more data points. Consider the scenario in which you are developing a mobile application to recognize chess pieces. Because the user's phone may not be completely perpendicular to the table where the chess board is set up, the chess pieces may seem to be rotated in either direction depending on how the phone is held. In this situation, random rotation may be an excellent option for simulating the appearance of numerous chess pieces without the need to precisely capture every possible angle of each piece.

It is also possible to counteract potential overfitting by using random rotation. Even in situations when the camera location is constant in relation to the subjects that your model is meeting, random rotation may enhance variety, preventing a model from retaining the training data that was used to train it (Solawetz, 2020).

3.3.2. Shifting augmentation

A horizontal shifting data augmentation implies that the image's pixels will be moved horizontally without affecting the image's overall size or proportions. As a result, some pixels of the image will be cut off and a new zone will be established in which new pixels of the image will be placed in order to reach the same image dimension as before. The width shift range and height shift range parameters of the Keras ImageDataGenerator class are used to move image pixels to the left, right, or top, bottom, respectively. These parameters, which we'll provide in the ImageDataGenerator function Object () [native code], can have floating-point values ranging from 0 to 1, and the value of these parameters affects how much the picture moves horizontally or vertically in a given direction. If we want to update a picture pixel, we can offer the pixel value as a pair of numbers between 0 and 1, rather than as a single integer between 0 and 1 (studytonight, 2021).

### 3.3.3. Shearing augmentation

Using the transformation matrix, you can do another bounding box transformation: shearing. Following is an example of the shearing effect (figure 11)

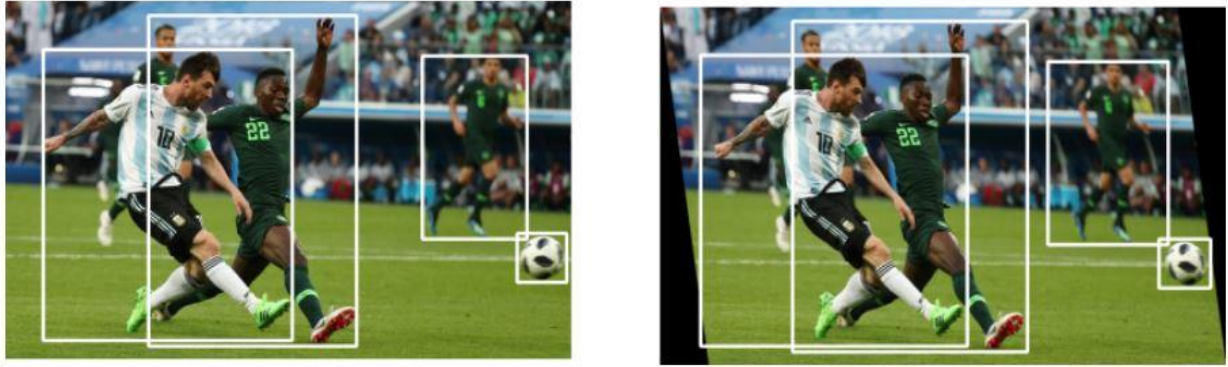

*Figure 11 shear effect*

Image slanting occurs due to shear transformation. Shear transformation differs from rotation in that it fixes one axis while stretching the picture at a certain angle called the shear angle. Because of this, the picture has a 'stretch' to it that is not shown in rotation (Sarin, 2021).

### 3.3.4. Zoom augmentation

Zoom Augmentation may be used to zoom in or out of the picture. There are two ways to use the ImageDataGenerator class:

The zoom range is [1-value, 1+value] if a single value is provided.

Lists of numbers may be used to determine a lower and upper bound for a particular number.

The picture may be zoomed in or out at random within the chosen range of zoom levels.

### 3.3.5. Flip augmentation

In a short amount of time, flipping an image (and the annotations on it) may improve a model's performance dramatically. For example, although machine learning models (such as convolutional neural networks) may memories a certain ordering of pixels that represents an item, our models may fail to recognize the same thing when it is mirrored across an image.

Take a moment to consider the multiple ways your face is understood whether you take a selfie or use the reverse lens on your camera. While one interpretation may be mirrored, the other may not, they are still your face. The phrase "flipping a picture" is used to denote this reversal of orientation.

We can give our deep learning model extra information to learn without having to go through the time-consuming process of obtaining and labelling new training data by making multiple copies of our pictures in various orientations.

3.3.6. Brightness change augmentation

When developing the model, the model's brightness has to be taken into account. Some of the photos may not have been captured in ideal lighting conditions. So, our model must be able to recognize the item even if it has the lowest possible resolution.

With our framework, we utilize an ImageDataGenerator class that takes a range of values and randomly selects one to set as the image's brightness.

3.3.7. Noise Injection augmentation

A large quantity of training data is not detrimental to deep learning neural networks. When the input data is tainted by noise, we have two new capabilities:

- Our deep neural network gets a lot more data to train on.
- On noisy data, we are able to train our neural network and it will perform well on noisy data as well.

When it comes to noise, machines have a harder time comprehending it than people do. In contrast to people, algorithms are incapable of ignoring or contextualizing noise. A neural network's capacity to produce correct predictions may be drastically altered by even the smallest, human-imperceptible pixel changes, which are known as "adversarial assaults." Image defects were examined by researchers at Arizona State University in an effort to better understand categorization methods. As far as classification accuracy is concerned, blurring and noise were shown to have a significant negative impact. Both pre-processing techniques and augmentation techniques may be used to introduce noise into a model's training set, allowing it to draw more accurate conclusions. As a rule of thumb, we don't want to introduce noise into our validation and testing sets, but we want to enhance the unpredictability of certain pictures for the purpose of training to prevent adversarial assaults and avoid overfitting (Nelson, 2020).

## 3.4. Method of data collection

The primary research approaches in this paper are a review of the literature and conceptual modelling. The main source of datasets for this project will be from KAGGLE.COM. the dataset used for testing the framework is called "Cats and Dogs Breeds Classification Oxford Dataset" which is published by "DR. AVICENNA" under the license "CC BY-SA 4.0".

Kaggle is A community platform for data scientists and machine learning aficionados, Kaggle enables them to cooperate and share their work. Working together with other data scientists, discovering and releasing datasets, using GPU notebooks, and competing against them are all alternatives available on Kaggle while tackling data science jobs. Google purchased this online platform in 2017 (created by Anthony Goldbloom and Jeremy Howard in 2010) in order to aid professionals and students in achieving their data science goals via the use of the platform's powerful tools and resources (Uslu, 2022).

## 3.5. Chapter summary

The expectations from the framework being built in this project is to reduce the time and effort spent on data management stage by creating an application that allows its users to perform data augmentation methods on their datasets without having to write any code in the process. The benefits of using data augmentation are many. Its not only good to solve the insufficient data problems but also increases the model accuracy even when the data are in hand.

# CHAPTER 4
## Implementation

### 4.1. Introduction

The framework was developed using python language which is a great choice to perform deep learning tasks. The application is using Keras library which is "a Python-based deep learning API that runs on top of TensorFlow, a machine learning framework. It was designed to facilitate rapid experimentation. The ability to quickly go from a concept to a result is essential for successful research (Keras Team, n.d.)".

The main class being used for image augmentation on this application is called Keras ImageDataGenerator. The Keras ImageDataGenerator class makes it simple to add additional data to your photos. Multiple augmentation methods, including standardization and rotation as well as a wide range of shifts and flips, are available. However, the primary advantage of utilizing the Keras ImageDataGenerator class is that it is meant to supplement real-time data. In other words, while your model is still learning, it is creating augmented pictures on the go. At each epoch, the ImageDataGenerator class provides the model with new photos. There are no new photos added to the original collection of photographs while using this method. One of the advantages of ImageDataGenerator is that it consumes less memory. In the absence of this class, all of the photos are loaded at once. However, when we use it, we save a lot of memory by loading photos in batches (Aniruddha, 2020).

The GUI was built using Tkinter package and the needed functions were created in order for the framework to work properly.

### 4.2. Coding

Figure 12 shows the packages used on this application and how it was being imported.

```
import os
import random
from tkinter import *
from tkinter import ttk

import numpy as np
from PIL import Image
from keras.preprocessing.image import ImageDataGenerator
from skimage import io
from ttkthemes import themed_tk as tkt
```

*Figure 12 the packages importing*

Figures (13-19) shows the Design implementation for the framework.

```
root = tkt.ThemedTk()
root.get_themes()
root.set_theme("yaru")
root.title('Image Augmentation')
root.geometry("1050x700+20+10")
root.config(background="MediumPurple1")
```

*Figure 13 GUI.1*

In Figure (13), we specified the dimensions of the UI which is 1050*700 and the overall theme for the application has been selected.

```python
label_title = ttk.Label(root, text="Image Augmentor", font=("Arial", 20, "bold"))
label_title.place(x=550, y=20, anchor="center")

DisplayLabel1 = ttk.Label(root, text="Write the path of the file/s here:", font=("Arial", 10))
DisplayLabel1.place(x=200, y=60)

txt_search = ttk.Entry(root, width=50)
txt_search.place(x=550, y=70, anchor="center")

DisplayLabel2 = ttk.Label(root, text="Where do you wish to save your augmented data:", font=("Arial", 10))
DisplayLabel2.place(x=100, y=120)

txt_search1 = ttk.Entry(root, width=50)
txt_search1.place(x=550, y=130, anchor="center")

btn_Clear = ttk.Button(root, text='clear', command=delete)
btn_Clear.place(x=720, y=54)

btn_Clear1 = ttk.Button(root, text='clear', command=delete1)
btn_Clear1.place(x=720, y=120)

label_title1 = ttk.Label(root, text="Select the augmentation technique", font=("Arial", 20, "bold"))
label_title1.place(x=550, y=170, anchor="center")

frame1 = ttk.Frame(root)
frame1.place(x=90, y=200)

var1 = IntVar()
```

*Figure 14 GUI.2*

```python
var1 = IntVar()
C1 = Checkbutton(frame1, text="Rotate (7)", width=100, anchor="w", variable=var1)
C1.pack(padx=10, pady=10)
C1.bind('<Leave>', onLeave)
C1.bind('<Enter>', onEnter)

var2 = IntVar()
C2 = Checkbutton(frame1, text="Shift (4)", width=100, anchor="w", variable=var2)
C2.pack(padx=10, pady=10)
C2.bind('<Leave>', onLeave2)
C2.bind('<Enter>', onEnter2)

var3 = IntVar()
C3 = Checkbutton(frame1, text="Shear (3)", width=100, anchor="w", variable=var3)
C3.pack(padx=10, pady=10)
C3.bind('<Leave>', onLeave3)
C3.bind('<Enter>', onEnter3)

var4 = IntVar()
C4 = Checkbutton(frame1, text="Zoom (4)", width=100, anchor="w", variable=var4)
C4.pack(padx=10, pady=10)
C4.bind('<Leave>', onLeave4)
C4.bind('<Enter>', onEnter4)

var5 = IntVar()
C5 = Checkbutton(frame1, text="Flip (4)", width=100, anchor="w", variable=var5)
C5.pack(padx=10, pady=10)
C5.bind('<Leave>', onLeave5)
```

*Figure 15 GUI.3*

```python
C5 = Checkbutton(frame1, text="Flip (4)", width=100, anchor="w", variable=var5)
C5.pack(padx=10, pady=10)
C5.bind('<Leave>', onLeave5)
C5.bind('<Enter>', onEnter5)

var6 = IntVar()
C6 = Checkbutton(frame1, text="Change Brightness (5)", width=100, anchor="w", variable=var6)
C6.pack(padx=10, pady=10)
C6.bind('<Leave>', onLeave6)
C6.bind('<Enter>', onEnter6)

var7 = IntVar()
C7 = Checkbutton(frame1, text="Add S&P noise (5)", width=100, anchor="w", variable=var7)
C7.pack(padx=10, pady=10)
C7.bind('<Leave>', onLeave7)
C7.bind('<Enter>', onEnter7)

var0 = IntVar()
C8 = Checkbutton(frame1, text="Select All", width=100, font=("Calibri", 12, "bold"), variable=var0, command=toggle_all)
C8.pack(padx=10, pady=10)
C8.bind('<Leave>', onLeave8)
C8.bind('<Enter>', onEnter8)

myButton = ttk.Button(frame1, text="Submit", command=lambda: Apply_btn())
myButton.pack(padx=10, pady=10)

chkList = [C1, C2, C3, C4, C5, C6, C7, C8]

root.mainloop()
```

*Figure 16 GUI.4*

Figures (14-15-16-) are showing the coding part for the UI including all the text areas, checkboxes, and buttons which will be triggering the assigned functions for each checkbox when the submit button is pressed.

```python
def onLeave(event):
    C1.config(bg='grey92')

def onLeave2(event):
    C2.config(bg='grey92')

def onLeave3(event):
    C3.config(bg='grey92')

def onLeave4(event):
    C4.config(bg='grey92')

def onLeave5(event):
    C5.config(bg='grey92')

def onLeave6(event):
    C6.config(bg='grey92')

def onLeave7(event):
    C7.config(bg='grey92')

def onLeave8(event):
    C8.config(bg='grey92')
```

*Figure 17 GUI.5*

```python
def onEnter(event):
    C1.config(bg='mint cream')

def onEnter2(event):
    C2.config(bg='mint cream')

def onEnter3(event):
    C3.config(bg='mint cream')

def onEnter4(event):
    C4.config(bg='mint cream')

def onEnter5(event):
    C5.config(bg='mint cream')

def onEnter6(event):
    C6.config(bg='mint cream')

def onEnter7(event):
    C7.config(bg='mint cream')

def onEnter8(event):
    C8.config(bg='mint cream')
```

*Figure 18 GUI.6*

In Figures (17-18), we defined the methods for hovering over each choice on the program to change the background color when it's hovered by the users.

```
def toggle_all():
    for cb in chkList:
        if var0.get():
            cb.select()
        else:
            cb.deselect()

def delete():
    txt_search.delete(0, 'end')
```

*Figure 19 GUI.7*

Figure (19) shows the toggle_all function which is responsible of selecting all the choices and ticking all the checkboxes when the "select all" checkbox is toggled. The other function (delete) will be triggered when the button clear is pressed which will clear all the written text on the text field for the path of the files.

The next figures (20 - 45) illustrate the implementation of the augmentation functions in the framework.

```python
def Rotating():
    datagen = ImageDataGenerator(
        rotation_range=180)
    
    dataset = []
    
    image_directory = txt_search.get()
    SIZE = 512
    dataset = []
    
    my_images = os.listdir(image_directory)
    for i, image_name in enumerate(my_images):
        if (image_name.split('.')[1] == 'jpg' or image_name.split('.')[1] == 'PNG'):
            image = io.imread(image_directory + image_name)
            image = Image.fromarray(image, 'RGB')
            image = image.resize((SIZE, SIZE))
            dataset.append(np.array(image))
    
    x = np.array(dataset)
    
    i = 0
    for batch in datagen.flow(x, batch_size=32, save_to_dir=txt_search1.get(), save_prefix='Rotated_img',
                              save_format='png'):
        i += 1
        if i >= 7:
            break
```

*Figure 20 Rotate Implementation*

When Rotating function is triggered, the program will perform rotation on the chosen images in 180 degrees randomly and then resize the new images to 512. After that the generated images will be given the proper labels and for each uploaded image, 7 new images will be generated.

```python
def Shear():
    datagen = ImageDataGenerator(
        shear_range=45, fill_mode='reflect')

    dataset = []

    image_directory = txt_search.get()
    SIZE = 512
    dataset = []

    my_images = os.listdir(image_directory)
    for i, image_name in enumerate(my_images):
        if (image_name.split('.')[1] == 'jpg' or image_name.split('.')[1] == 'PNG'):
            image = io.imread(image_directory + image_name)
            image = Image.fromarray(image, 'RGB')
            image = image.resize((SIZE, SIZE))
            dataset.append(np.array(image))

    x = np.array(dataset)

    i = 0
    for batch in datagen.flow(x, batch_size=32, save_to_dir=txt_search1.get(), save_prefix='Sheared_img',
                              save_format='png'):
        i += 1
        if i >= 3:
            break
```

*Figure 21 Shear implementation*

When Shear function is triggered, the program will perform a shear on the chosen images in a range of 45 and the blank side will be filled with reflection of the original image. The new images will be resized to 512. After that the generated images will be given the proper labels and for each uploaded image, 3 new images will be generated.

```python
def Shift():
    datagen = ImageDataGenerator(
        width_shift_range=0.2, height_shift_range=0.2, fill_mode='reflect')

    dataset = []

    image_directory = txt_search.get()
    SIZE = 512
    dataset = []

    my_images = os.listdir(image_directory)
    for i, image_name in enumerate(my_images):
        if (image_name.split('.')[1] == 'jpg' or image_name.split('.')[1] == 'PNG'):
            image = io.imread(image_directory + image_name)
            image = Image.fromarray(image, 'RGB')
            image = image.resize((SIZE, SIZE))
            dataset.append(np.array(image))

    x = np.array(dataset)

    i = 0
    for batch in datagen.flow(x, batch_size=32, save_to_dir=txt_search1.get(), save_prefix='Shifted_img',
                              save_format='png'):
        i += 1
        if i >= 4:
            break
```

*Figure 22 Shift implementation*

When Shift function is triggered, the program will perform a shifting on the chosen images in a range of 0.2 on the width and the height and the blank side will be filled with reflection of the original image. The new images will be resized to 512. After that the generated images will be given the proper labels and for each uploaded image, 4 new images will be generated.

```python
def Zoom():
    datagen = ImageDataGenerator(
        zoom_range=0.5, fill_mode='reflect')

    dataset = []

    image_directory = txt_search.get()
    SIZE = 512
    dataset = []

    my_images = os.listdir(image_directory)
    for i, image_name in enumerate(my_images):
        if (image_name.split('.')[1] == 'jpg' or image_name.split('.')[1] == 'PNG'):
            image = io.imread(image_directory + image_name)
            image = Image.fromarray(image, 'RGB')
            image = image.resize((SIZE, SIZE))
            dataset.append(np.array(image))

    x = np.array(dataset)

    i = 0
    for batch in datagen.flow(x, batch_size=32, save_to_dir=txt_search1.get(), save_prefix='Zoomed_img',
                              save_format='png'):
        i += 1
        if i >= 4:
            break
```

*Figure 23 Zoom implementation*

When Zoom function is triggered, the program will perform a zooming in and out of the chosen images in a range of 50% and the blank side will be filled with reflection of the original image. The new images will be resized to 512. After that the generated images will be given the proper labels and for each uploaded image, 4 new images will be generated.

```python
def Flip():
    datagen = ImageDataGenerator(
        horizontal_flip=True, vertical_flip=True, fill_mode='reflect')

    dataset = []

    image_directory = txt_search.get()
    SIZE = 512
    dataset = []

    my_images = os.listdir(image_directory)
    for i, image_name in enumerate(my_images):
        if (image_name.split('.')[1] == 'jpg' or image_name.split('.')[1] == 'PNG'):
            image = io.imread(image_directory + image_name)
            image = Image.fromarray(image, 'RGB')
            image = image.resize((SIZE, SIZE))
            dataset.append(np.array(image))

    x = np.array(dataset)

    i = 0
    for batch in datagen.flow(x, batch_size=32, save_to_dir=txt_search1.get(), save_prefix='Flipped_img',
                              save_format='png'):
        i += 1
        if i >= 4:
            break
```

*Figure 24 Flip implementation*

When Flip function is triggered, the program will perform a flipping on the chosen images horizontally and vertically and the blank side will be filled with reflection of the original image. The new images will be resized to 512. After that the generated images will be given the proper labels and for each uploaded image, 4 new images will be generated.

```python
def ChangeBrightness():
    datagen = ImageDataGenerator(
        brightness_range=[0.2, 2.1])

    dataset = []

    image_directory = txt_search.get()
    SIZE = 512
    dataset = []

    my_images = os.listdir(image_directory)
    for i, image_name in enumerate(my_images):
        if (image_name.split('.')[1] == 'jpg' or image_name.split('.')[1] == 'PNG'):
            image = io.imread(image_directory + image_name)
            image = Image.fromarray(image, 'RGB')
            image = image.resize((SIZE, SIZE))
            dataset.append(np.array(image))

    x = np.array(dataset)

    i = 0
    for batch in datagen.flow(x, batch_size=32, save_to_dir=txt_search1.get(), save_prefix='Brightness_Changed_img',
                              save_format='png'):
        i += 1
        if i >= 5:
            break
```

*Figure 25 Brightness implementation*

When ChangeBrightness function is triggered, the program will perform a random change in the brightness of the chosen images in a range between 0.2 and 2.1 where 0 is the absolute dark. The new images will be resized to 512. After that the generated images will be given the proper labels and for each uploaded image, 5 new images will be generated.

```python
def add_noise(img):
    '''Add random noise to an image'''
    VARIABILITY = 50
    deviation = VARIABILITY * random.random()
    noise = np.random.normal(0, deviation, img.shape)
    img += noise
    np.clip(img, 0., 255.)
    return img
```

*Figure 26 Noise implementation.1*

```python
def AddNoise():
    datagen = ImageDataGenerator(
        preprocessing_function=add_noise,
    )

    dataset = []

    image_directory = txt_search.get()
    SIZE = 512
    dataset = []

    my_images = os.listdir(image_directory)
    for i, image_name in enumerate(my_images):
        if (image_name.split('.')[1] == 'jpg' or image_name.split('.')[1] == 'PNG'):
            image = io.imread(image_directory + image_name)
            image = Image.fromarray(image, 'RGB')
            image = image.resize((SIZE, SIZE))
            dataset.append(np.array(image))

    x = np.array(dataset)

    i = 0
    for batch in datagen.flow(x, batch_size=32, save_to_dir=txt_search1.get(), save_prefix='Noise_injected_img',
                              save_format='png'):
        i += 1
        if i >= 5:
            break
```

*Figure 27 Noise implementation.2*

When AddNoise function is triggered, the program will add salt & pepper noise to the chosen images. The new images will be resized to 512. After that the generated images will be given the proper labels and for each uploaded image, 5 new images will be generated.

4.3.    Testing

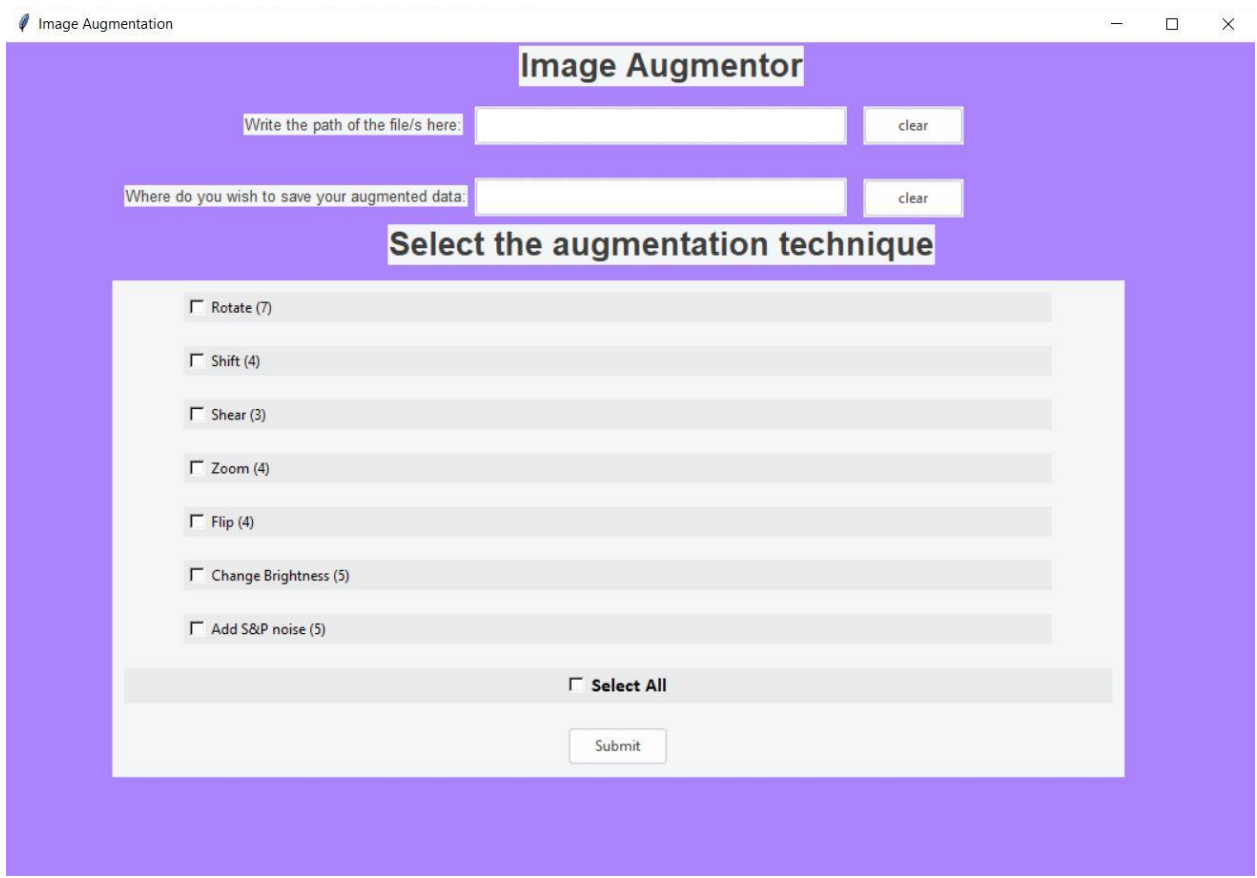

*Figure 28 Testing the GUI*

Figure 28 shows the UI of the application after running the code in PyCharm compiler.

For the testing, each and every function and button will be tested in different scenarios to ensure that the application is working as it is intended to be used.

## 4.4. Summary

The coding was done in python language. There were several packages and libraries used on this application along with plain programming to complete all the functionalities in the application.

The code was tested in PyCharm compiler and the UI has been initiated successfully. All the checkboxes and buttons are functioning properly. The application is supposed to perform all kind of augmentations listed on the UI which will be illustrated and discussed on next chapter (Results and analysis).

# CHAPTER 5
# RESULTS AND ANALYSIS

## 5.1. Introduction

This chapter will mainly focus on the analysis of the results of the research and the implementation of the framework.

## 5.2. Results & analysis

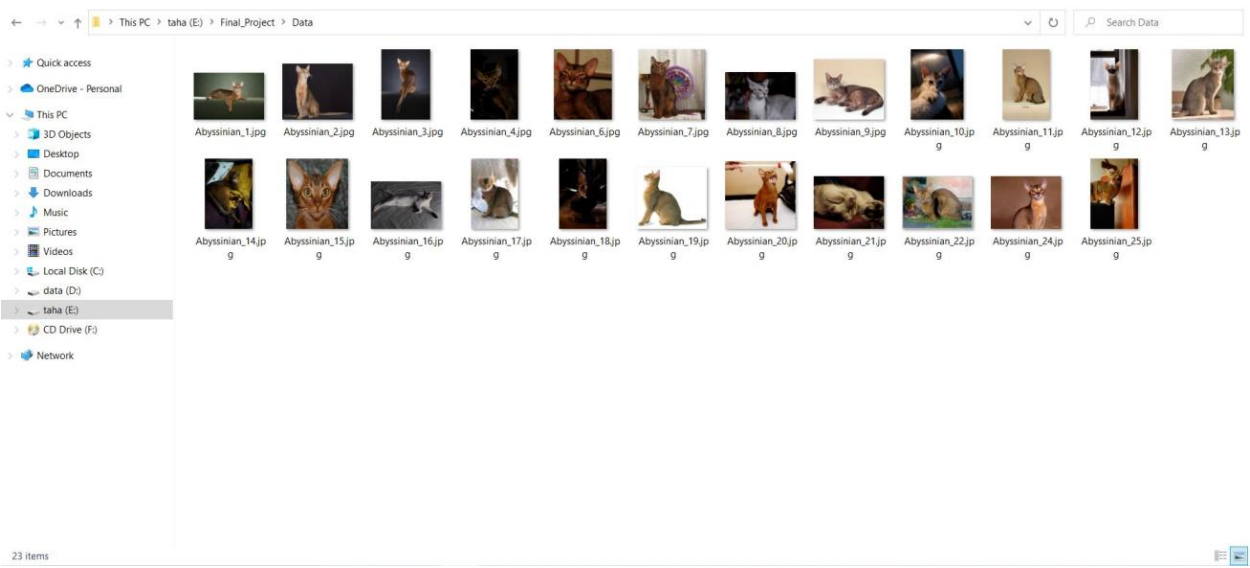

*Figure 29 Original dataset*

Figure (29) shows the original dataset which was downloaded from "KAGGLE.COM- Cats and Dogs Breeds Classification Oxford Dataset "

### 5.2.1. Rotate augmentation results

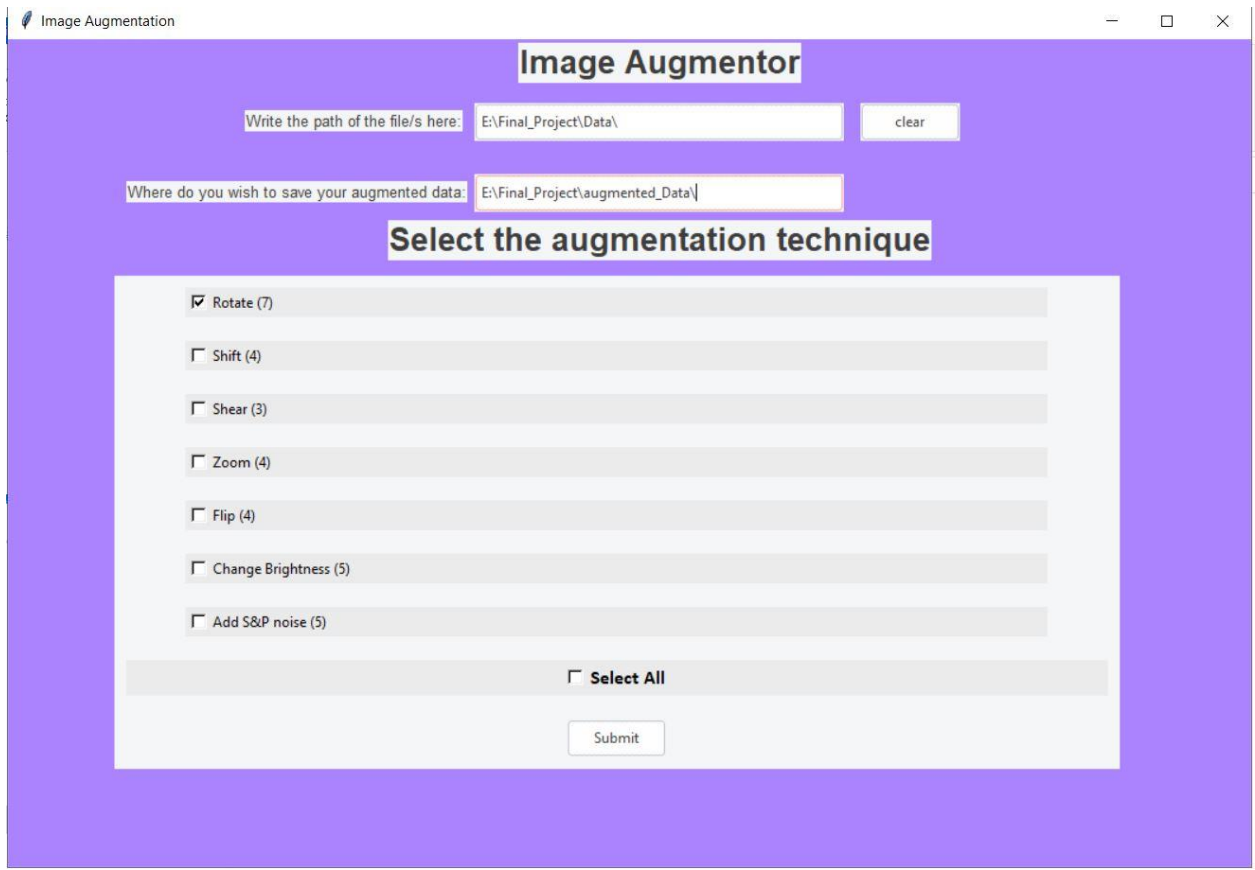

*Figure 30 Rotate testing*

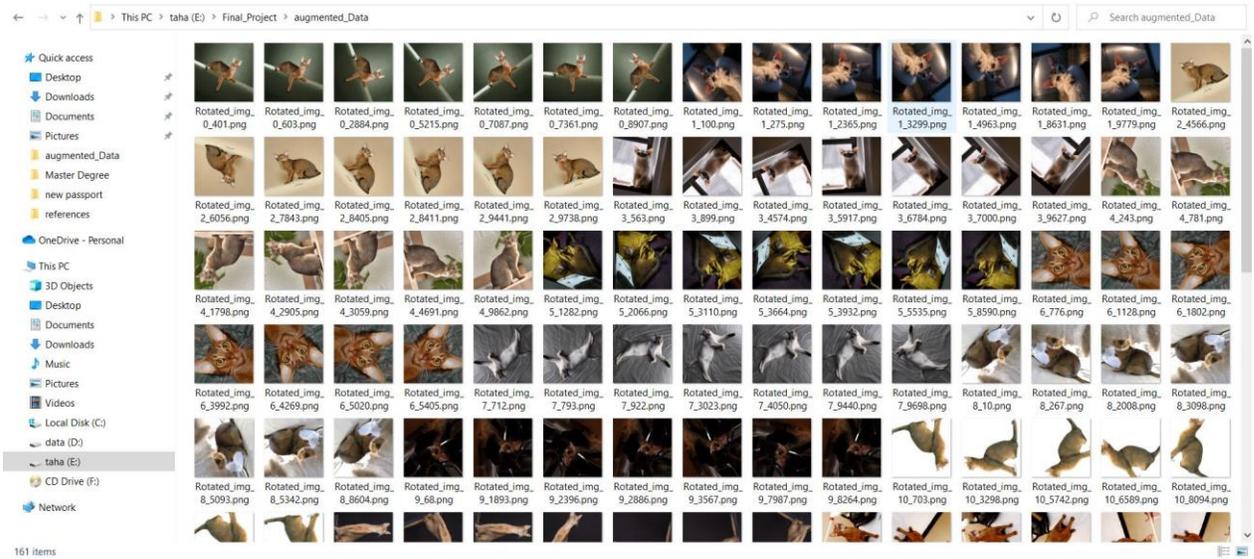

*Figure 31 After performing rotate*

The results of rotating augmentation are shown in figure 31 where 161 new images were generated from the original dataset and those generated images are given the proper labels.

### 5.2.2. Shift augmentation results

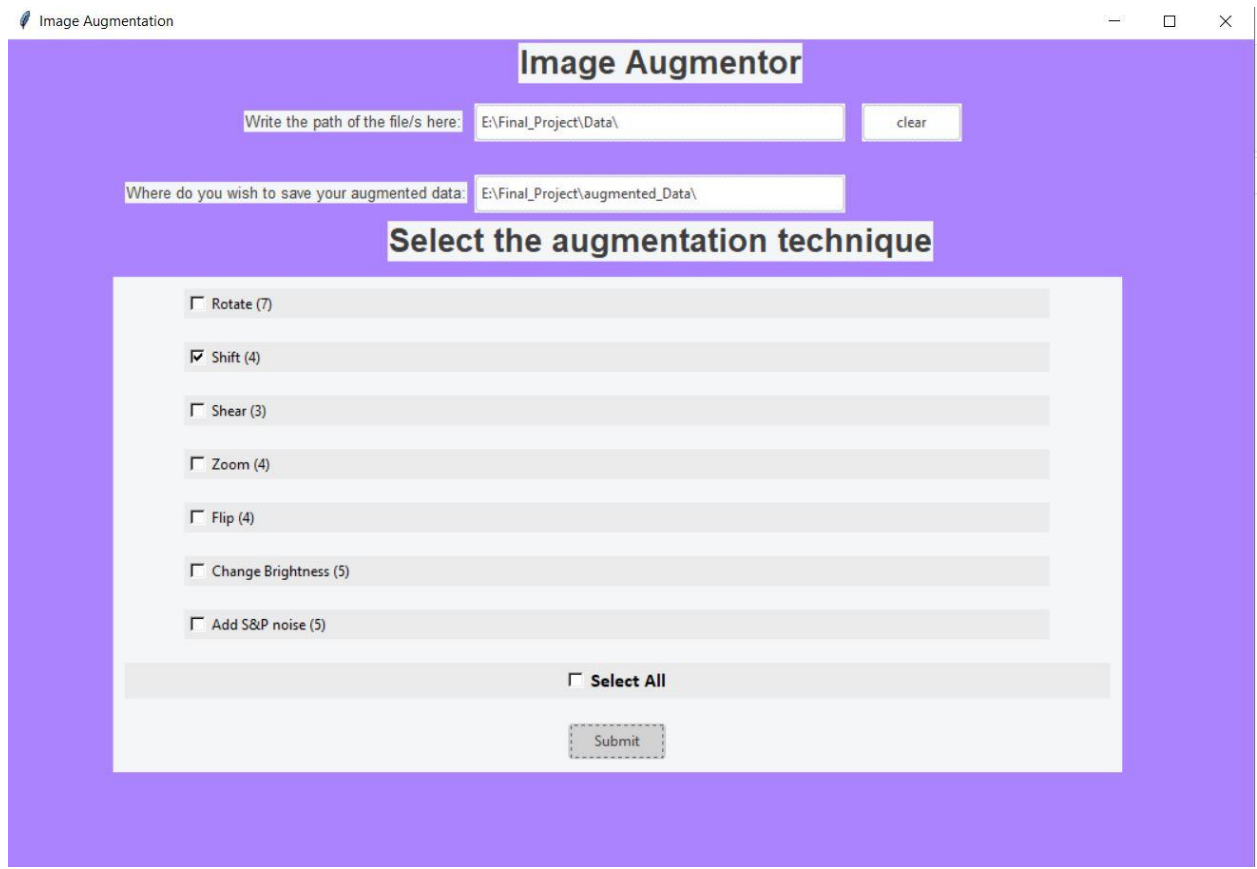

*Figure 32 shift testing*

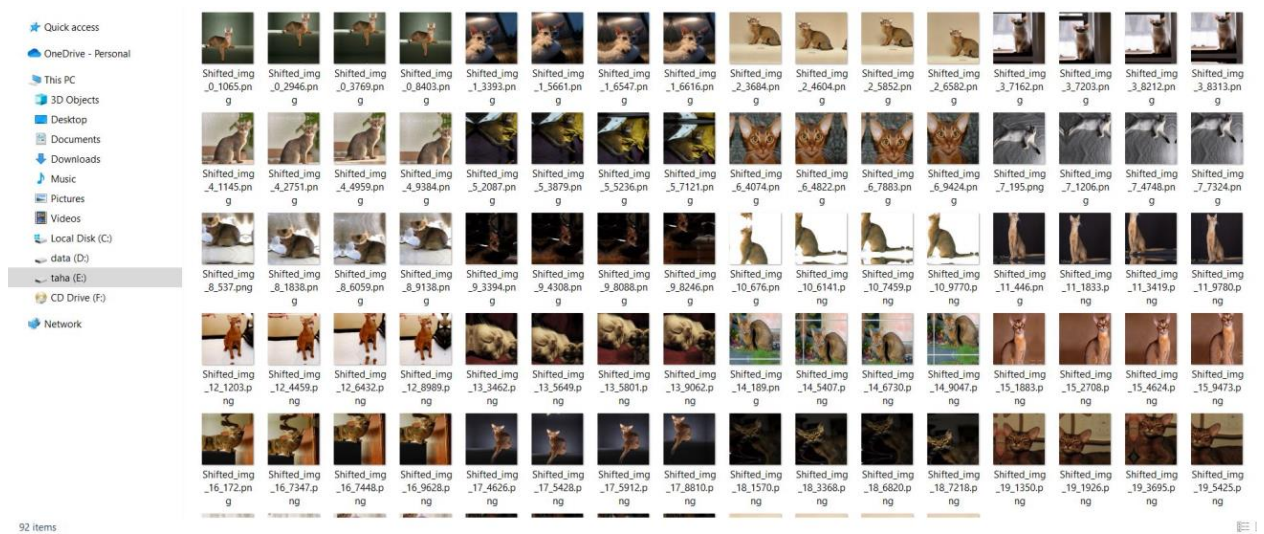

*Figure 33 After shifting*

The results of shift augmentation are shown in figure 33 where 92 new images were generated from the original dataset and those generated images are given the proper labels.

### 5.2.3. Shear augmentation results

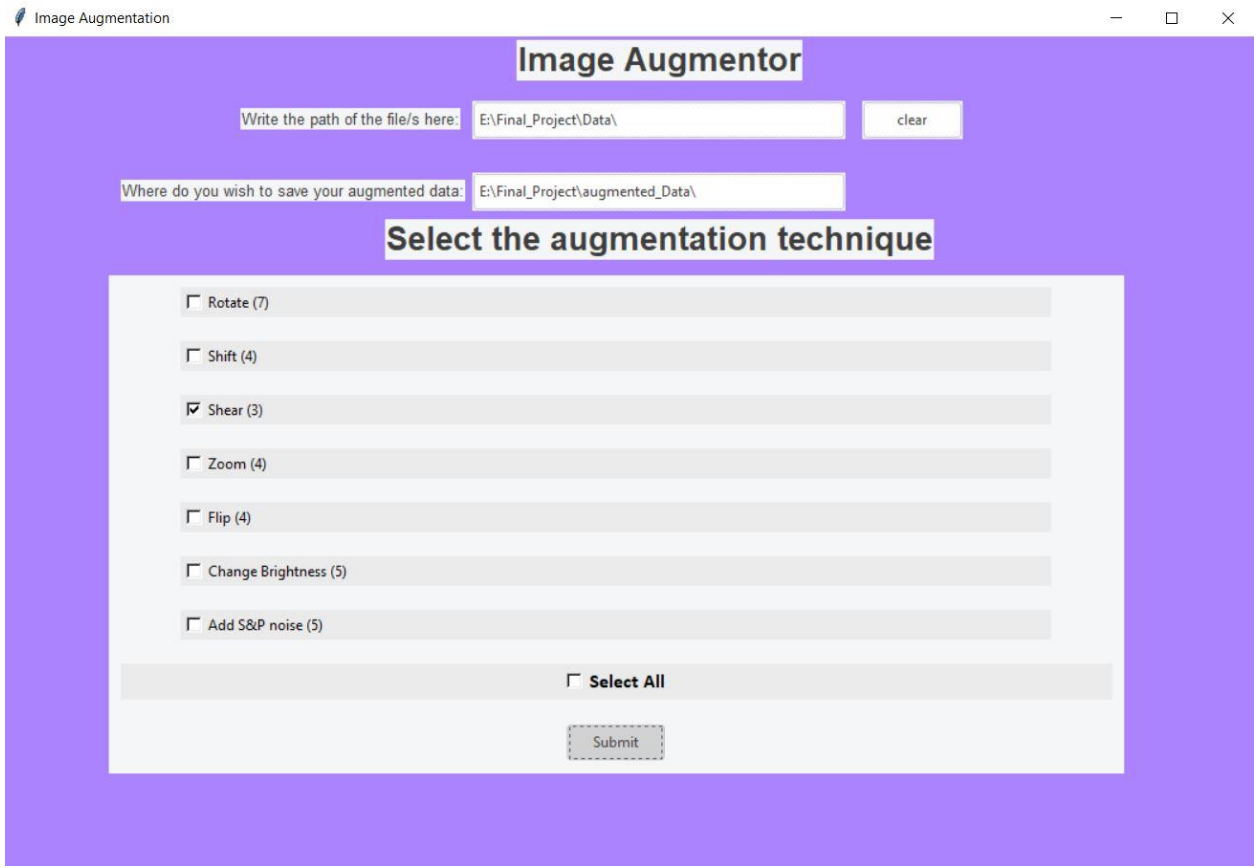

*Figure 34 Shear testing*

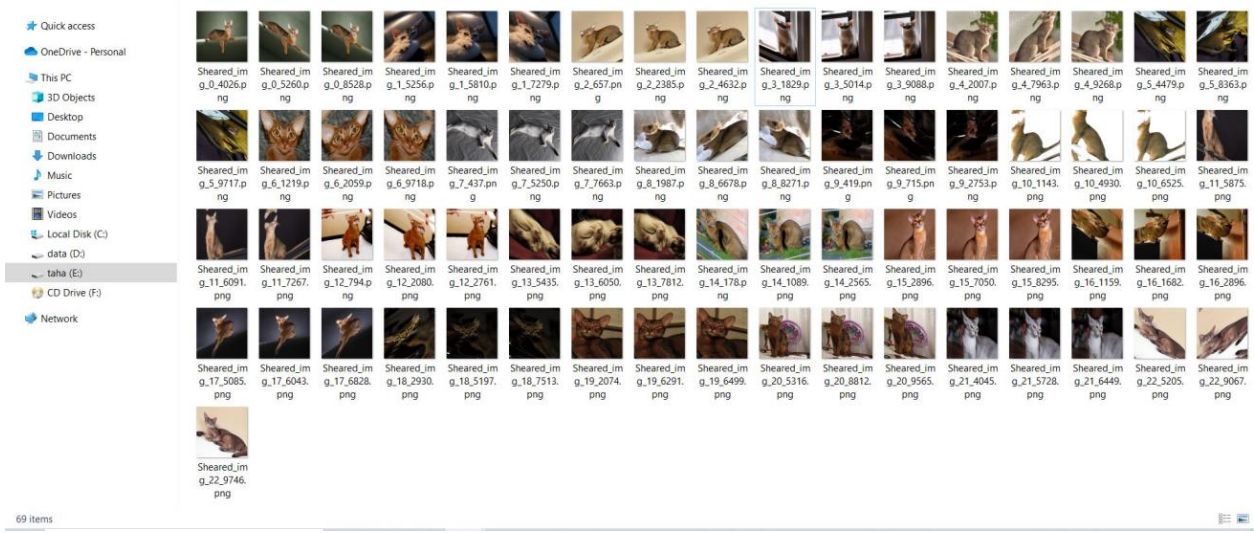

*Figure 35 After shear*

The results of shear augmentation is shown in figure 35 where 69 new images were generated from the original dataset and those generated images are given the proper labels.

### 5.2.4. Zoom augmentation results

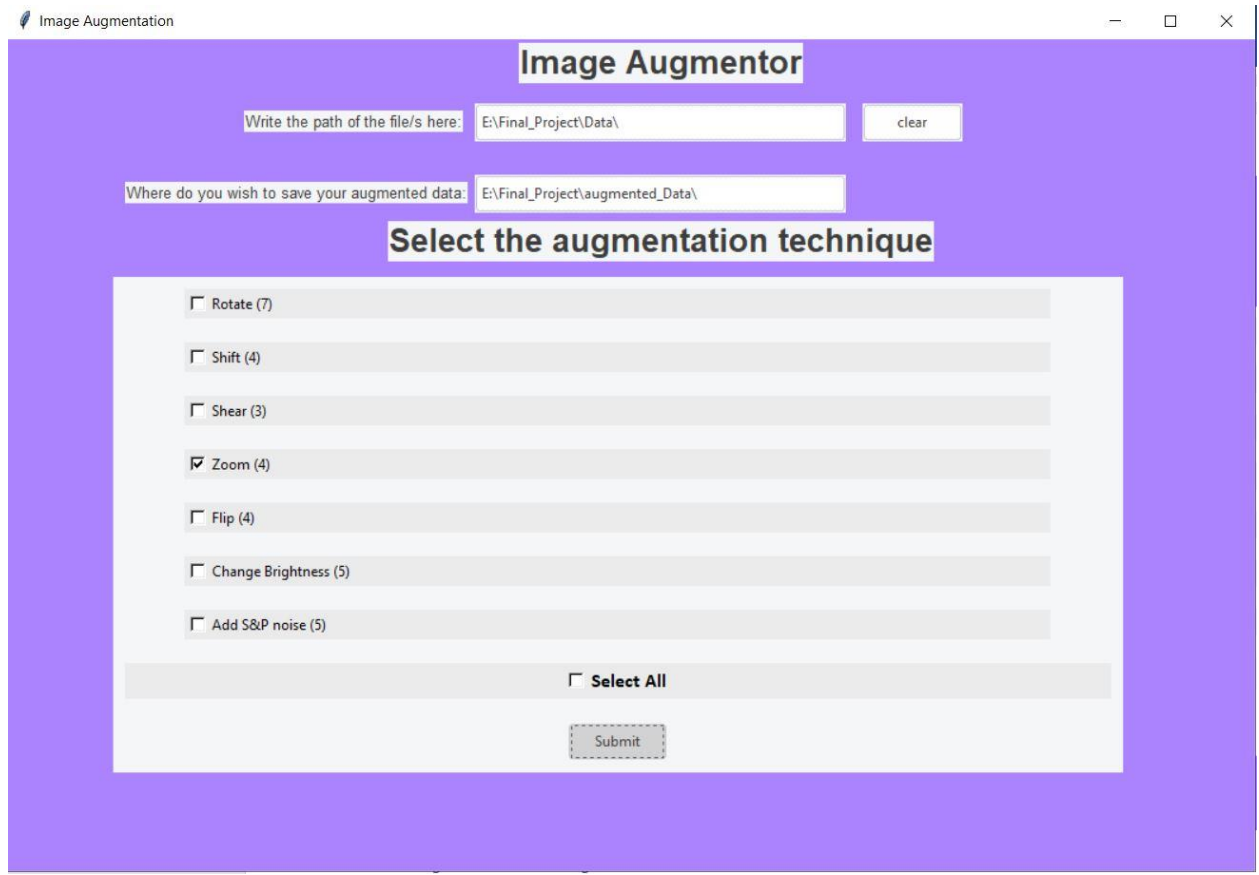

*Figure 36 zoom testing*

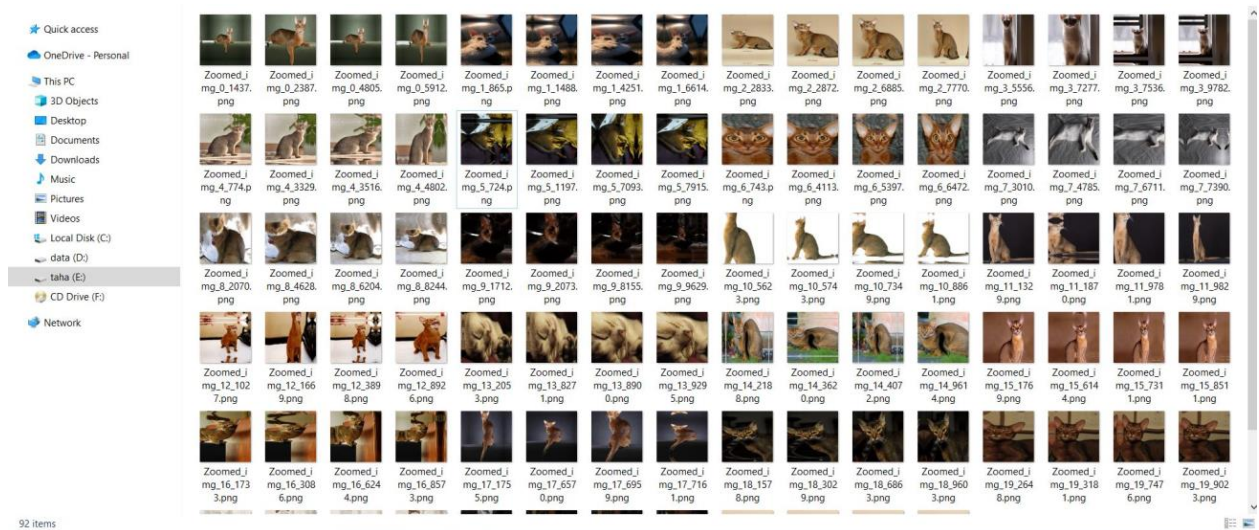

*Figure 37 After zooming*

The results of zoom augmentation are shown in figure 37 where 92 new images were generated from the original dataset and those generated images are given the proper labels.

### 5.2.5. Flip augmentation results

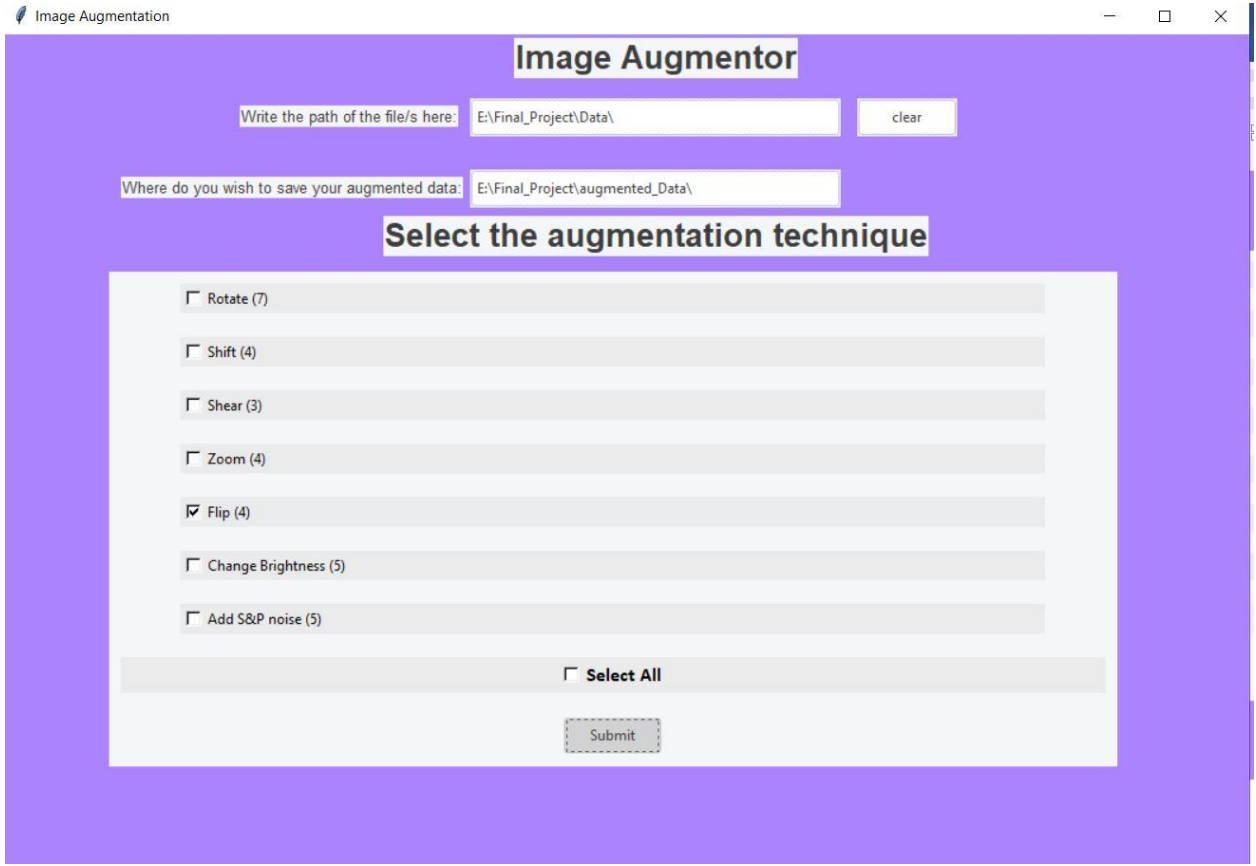

*Figure 38 Flip testing*

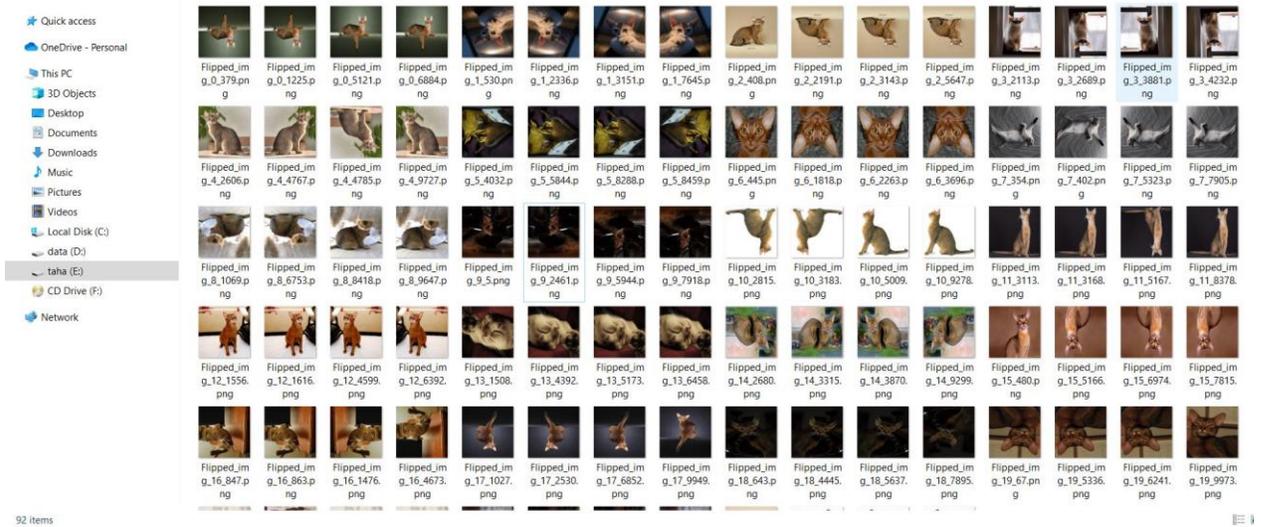

*Figure 39 after flipping*

The results of flip augmentation are shown in figure 33 where 92 new images were generated from the original dataset and those generated images are given the proper labels.

### 5.2.6. Brightness augmentation results

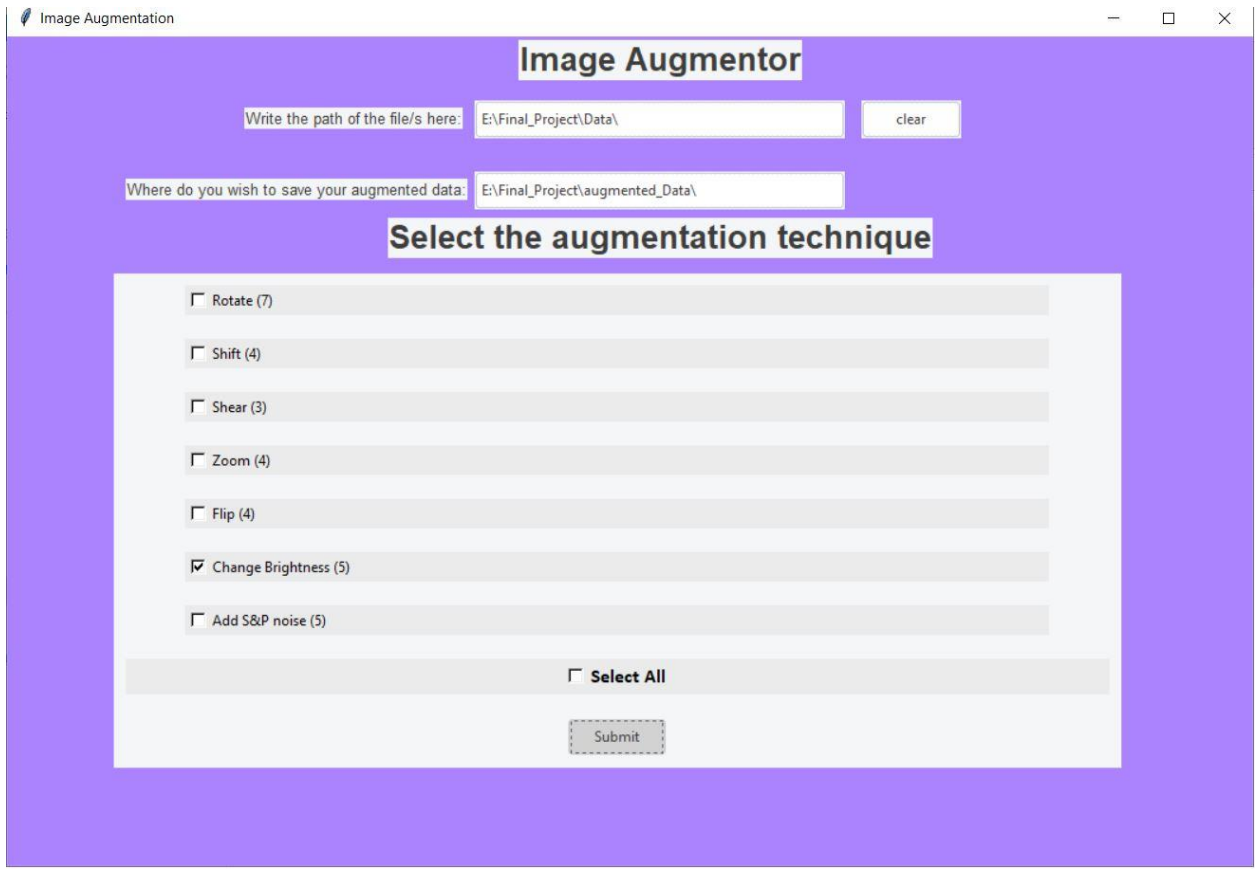

*Figure 40 Brightness change testing*

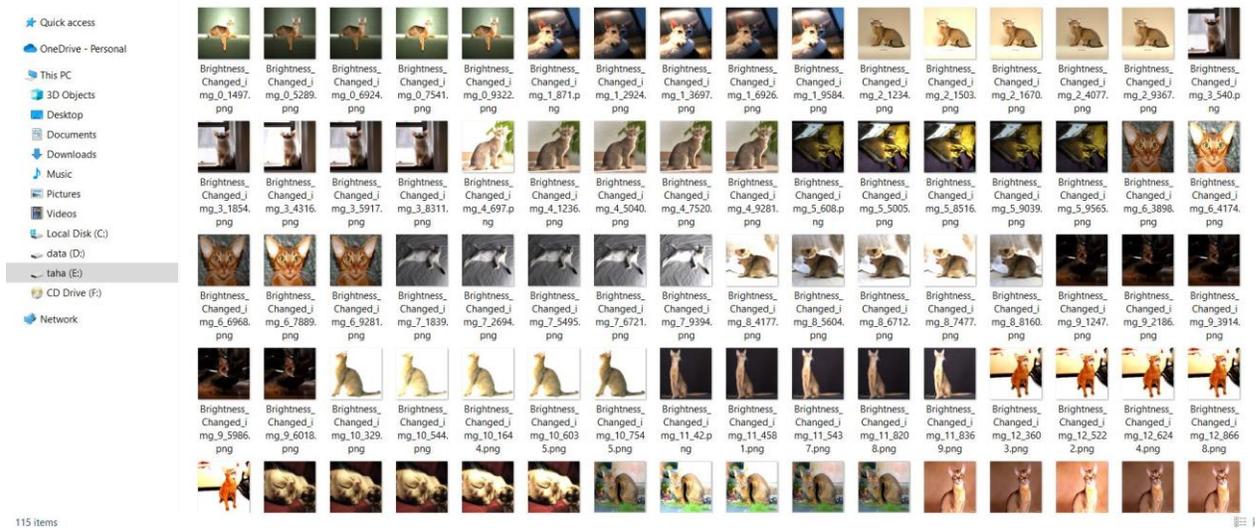

*Figure 41 after brightness changed*

The results of brightness augmentation are shown in figure 41 where 115 new images were generated from the original dataset and those generated images are given the proper labels.

### 5.2.7. Noise injection augmentation results

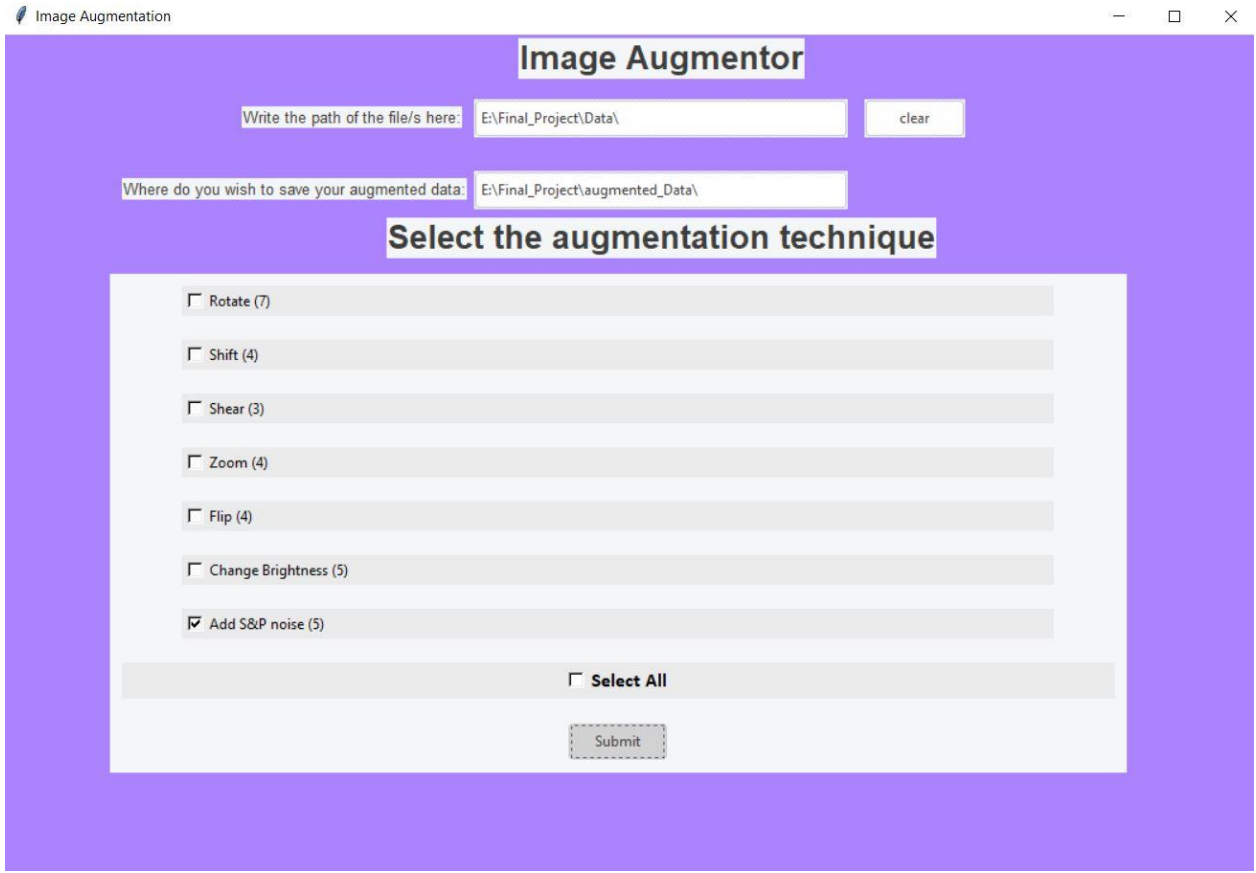

*Figure 42 Salt & Pepper noise testing*

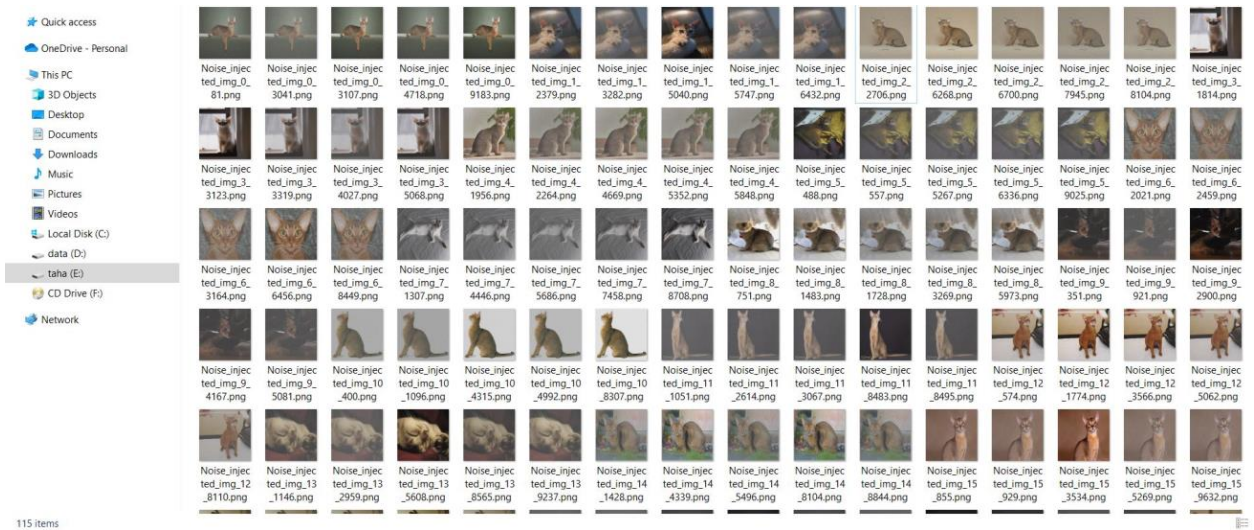

*Figure 43 After S&P noise injected*

The results of noise injection augmentation are shown in figure 43 where 115 new images were generated from the original dataset and those generated images are given the proper labels.

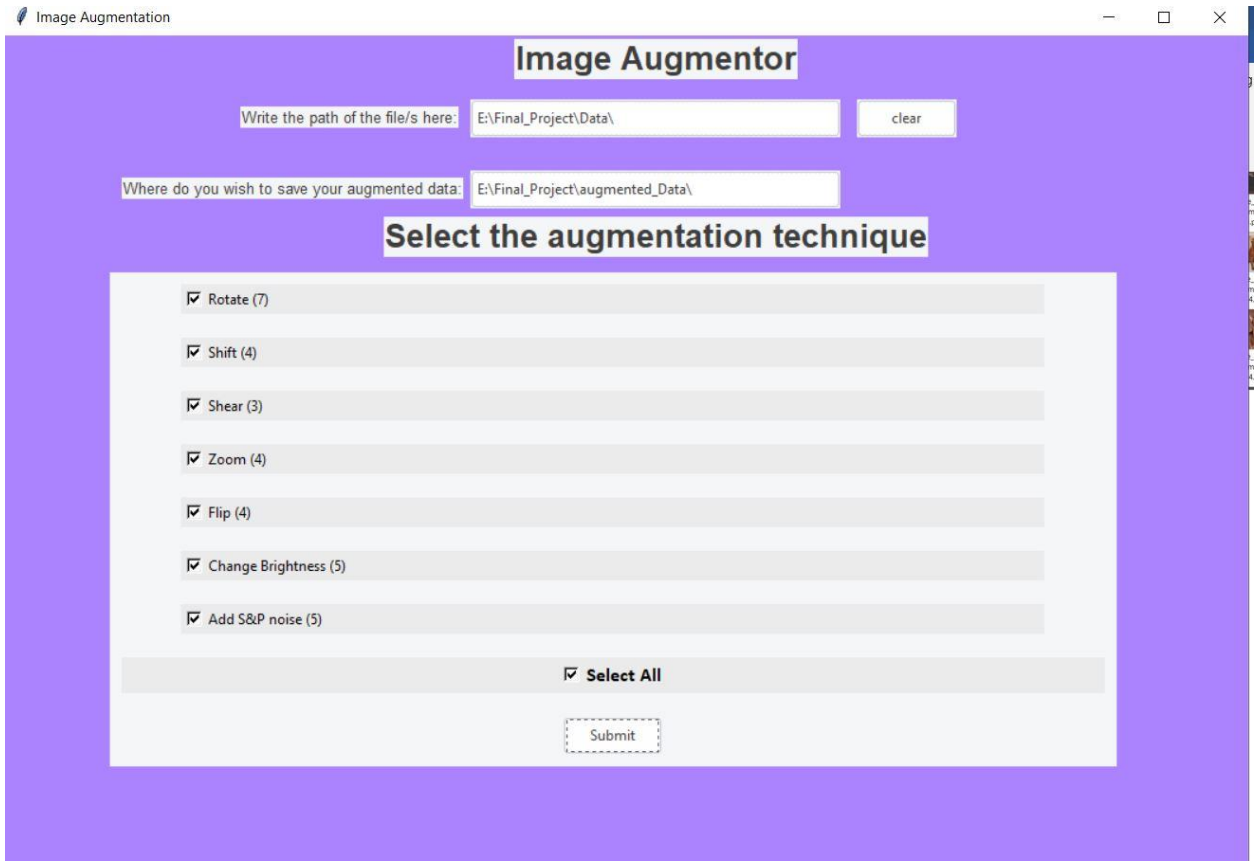

*Figure 44 select all testing*

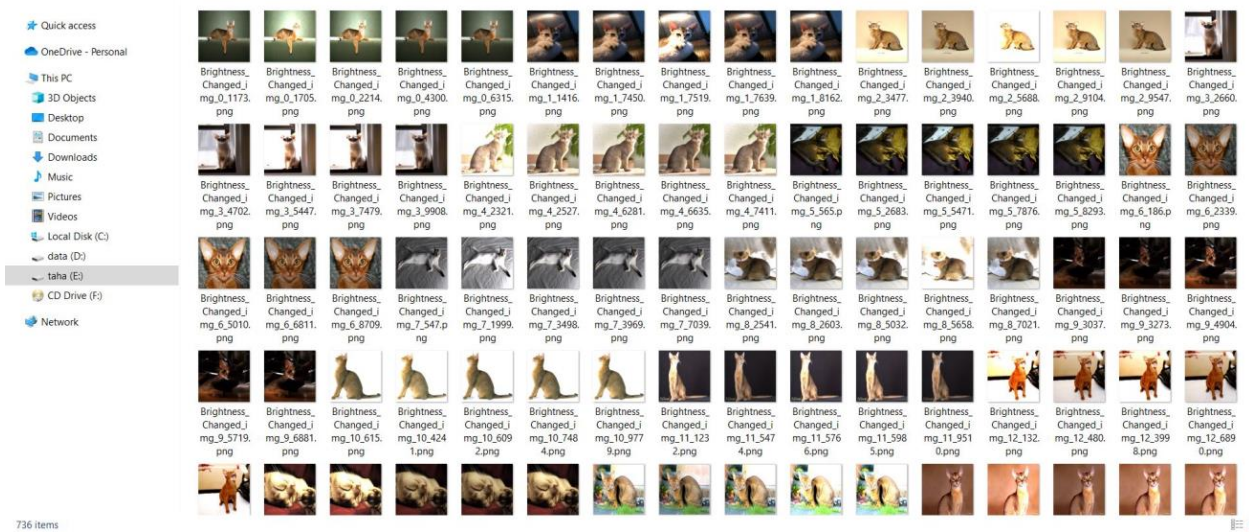

*Figure 45 after performing all augmentations together*

The select all checkbox was tested and all the augmentations were performed at the same time where 736 new images were generated in total.

5.3. Evaluation of the Proposed Method

The utilization of the built framework which can perform several image augmentations (Rotating images – Shifting images – Shearing images – Zooming in and out images – Flipping images – Changing the brightness of the images – Injecting salt and pepper noise to images) which has proven their effectiveness on the machine learning applications where data are not sufficient enough (Shorten & Khoshgoftaar, 2019), will help the researchers and developers on performing the mentioned types of image augmentations during data management stage of machine learning development without having to write any code which can save big amount of time and effort. The built framework will also assign new labels to the generated images which can specify what kind of augmentation have been done on these images like rotated images will have a label like "Rotated_img_12_7424". Its possible to assign the class name of images to the generated ones like in the case of generating new images for the class dogs, the images can have a label like rotated_dog_images to save even more time that is usually required to augment and label the augmented data.

The issue with using tiny datasets to train models is that the resulting models do not generalize the data effectively when applied to the validation and test sets. As a result, these models have the issue of being too accommodating (Perez & Wang, 2017). The data augmentation utilization with the framework will help on solving the overfitting problems in the data preparation.

# CHAPTER 6
# CONCLUSION AND FUTURE RECOMMENDATION

## 6.1. Introduction

This chapter will concentrate mostly on the project's conclusion, which will include a summary of the findings. In addition, the study's limitations and proposals for improving the validity of the results will be discussed. Additional areas of study for future advanced research will be discussed in this chapter.

## 6.2. Summary

The development life cycle for a machine learning application is consisting of multiple stage where each and every stage depends on the prior stages to be successful in order for it to succeed as well. One the most important stage of ML development is data management stage where most of the data-related activities are being conducted. Since the ML applications depends heavily on data, its crucial to ensure that this stage was done in the best possible way. Many issues are being encountered during DM stage and therefore, its essential to solve each and every problem faced to have a well-developed machine learning application.

Data scarcity is a serious problem being faced daily when developing ML application. AS many sectors keep their data confidential, its getting harder for developers and researchers to obtain the amount of data they need to build am accurate machine learning model.

Because of this issue and many others when it comes to DM stage, the researchers have been focusing heavily since the beginning of ML applications on finding solutions and developing advancements that can help overcoming these issues. One of the most well-knows techniques/solutions for data, is called data augmentation. Data augmentation is a technique used to generate new data from the existing data to overcome the lack of data. Not only that but also to improve the quality and accuracy of the model as data augmentation has proven over the years that its not only a fantastic solution for the lack of data but its also great solution to improve the accuracy and quality of the model even when there are available data at hand.

With the help of Software engineering skills and tools, we were able to build a framework that is completely coding free for its users who want to utilize data augmentation techniques without writing a single line of code. This framework allows you to perform several types of data augmentation on your dataset to improve its quality and quantity for a better machine learning application. The framework was build using python programming language along with many libraries that help in constructing a data augmenter for researchers and developers. The framework was tested properly and each and every function and type of augmentation is working as intended.

6.3. Limitation of the study

The time was a serious limitation on this project as we had only 3 months to complete it. Due to the lack of time, the framework was limited to the image type of data so that it only can read and generate images as the type of data. Another limitation on this project was the fact that it was done during the pandemic of Covid-19 which kept us from working physically with the best possible equipments for this project. This project was completely done from home which affected the kind of tools to be used on this project to keep it at the level where the personal computers can handle so only the tools that can be utilized on the available computers were used.

6.4. Future Enhancements

This project can be enhanced in many aspects. The first aspect would be to have a proper implementation for all kind of augmentations available at the moment which was not all included because of the lack of time. One of the most important technique that can enhance this project is GAN (Generative Adversarial Network).

Using Generative Adversarial Networks (GANs), you may generate new data samples. The original dataset's feature distribution is mimicked by GANs, which use latent space noise to generate new pictures. Within the design of a GAN, there are two networks vying to reach Nash Equilibrium (More on Game Theory and what Nash Equilibrium is can be found here). Using the DCGAN and WGAN, two competing models of Convolutional Neural Networks, the GAN networks were modelled for this study. As with DCGANs, WGANs employ the Wasserstein-1 Distance as their loss function, making the discriminator a "critic" that forecasts how probable a sample is from the original data. It has been used to create fresh data and to enhance the performance of CNNs using Generative Adversarial Networks (GANs). In spite of this, data augmentation strategies for training GANs have been understudied in comparison to CNNs. An entirely new GAN architecture has been proposed in this study to enhance chest X-rays for the identification of pneumonia and COVID-19 utilizing generative models. (Motamed et al., 2021).

The "Discriminator" in a GAN is solely responsible for determining whether a sample is fake or authentic. True samples originate from the actual dataset, while false samples are generated by a generator. The "Generator," as implied by its name, is designed to generate "false" data for the discriminator by accepting random noise from a latent space as an input. When the discriminator sees a picture generated by the generator, it assumes that it is real. Using backpropagation, the generator will gradually learn to generate samples that closely match the original dataset's distribution in terms of weights and biases (McCloskey, 2022).

different color spaces in clustering based image segmentation. Communications in Computer and Information Science 81 PART 2 532–541. https://doi.org/10.1007/978-3-642-14058-7_55

Kang G. Dong X. Zheng L. & Yang Y. (2017). PatchShuffle Regularization. https://doi.org/10.48550/arxiv.1707.07103

Karras T. Aila T. Laine S. & Lehtinen J. (2018). PROGRESSIVE GROWING OF GANS FOR IMPROVED QUALITY STABILITY AND VARIATION. ICLR. https://youtu.be/G06dEcZ-QTg.

Kamalakannan, M., & Khamis, T. (2022). Assessing tools to analyze the techniques and mechanism for network risk minimization. Journal of Applied Technology and Innovation, 6(1), 14. https://www.researchgate.net/publication/357554036_Assessing_tools_to_analyze_the_techniques_and_mechanism_for_network_risk_minimization#fullTextFileContent

Khalid, M., Mokayed, H., Yusof, R., & Ono, O. (2009). Online signature verification with neural networks classifier and fuzzy inference. Proceedings - 2009 3rd Asia International Conference on Modelling and Simulation, AMS 2009, 236–241. https://doi.org/10.1109/AMS.2009.23

Khalid, M., Yusof, R., & Mokayed, H. (2011). Fusion of Multi-Classifiers for online signature verification using fuzzy Logic Inference. International Journal of Innovative Computing, Information and Control, 7(5 B), 2709–2726.

Khamis, A. A., Idris, A., Abdellatif, A., Mohd Rom, N. A., Khamis, T., Ab Karim, M. S., Janasekaran, S., & Abd Rashid, R. Bin. (2023). Development and Performance Evaluation of an IoT-Integrated Breath Analyzer. International Journal of Environmental Research and Public Health, 20(2). https://doi.org/10.3390/IJERPH20021319
71